\documentclass{article}

\pdfoutput=1
\usepackage{PRIMEarxiv}

\usepackage[utf8]{inputenc} 
\usepackage[T1]{fontenc}    
\usepackage{url}            
\usepackage{booktabs}       
\usepackage{amsfonts}       
\usepackage{nicefrac}       
\usepackage{microtype}      
\usepackage{lipsum}
\usepackage{fancyhdr}       
\usepackage{graphicx}       
\graphicspath{{media/}}     
\usepackage{natbib}
\setcitestyle{authoryear} 
\usepackage[figuresright]{rotating}
\usepackage{amsmath}
\usepackage{amsfonts}

\usepackage[T1]{fontenc}
\usepackage{newpxmath}
\usepackage{hyperref}
\hypersetup{
    colorlinks,
    linkcolor={blue},
    citecolor={blue},
    urlcolor={black}
}

\usepackage{threeparttable}

\newtheorem{assumption}{Assumption}

\newcommand{\indep}{\perp \!\!\! \perp}
\usepackage{mathtools}
\usepackage{multirow} 

\title{On the Proportional Principal Stratum Hazards Model}

\author{
  Jiren Sun \\
  Department of Biostatistics and Medical Informatics \\
  University of Wisconsin--Madison \\
  Madison, WI, 53726 \\
\texttt{jiren.sun@wisc.edu} \\
   \And
   Thomas D. Cook \\
  Department of Biostatistics and Medical Informatics \\
  University of Wisconsin--Madison \\
  Madison, WI, 53726 \\
   \texttt{cook@biostat.wisc.edu} \\
}

\begin{document}
\maketitle

\begin{abstract}
In clinical trials involving both mortality and morbidity, an active treatment can influence the observed risk of the first non-fatal event either directly, through its effect on the underlying non-fatal event process, or indirectly, through its effect on the death process, or both. Discerning the direct effect of treatment on the underlying first non-fatal event process holds clinical interest. However, with the competing risk of death, the Cox proportional hazards model that treats death as non-informative censoring and evaluates treatment effects on time to the first non-fatal event provides an estimate of the cause-specific hazard ratio, which may not correspond to the direct effect. To obtain the direct effect on the underlying first non-fatal event process, within the principal stratification framework, we define the principal stratum hazard and introduce the Proportional Principal Stratum Hazards model. This model estimates the principal stratum hazard ratio, which reflects the direct effect on the underlying first non-fatal event process in the presence of death and simplifies to the hazard ratio in the absence of death. The principal stratum membership is identified probabilistically using the shared frailty model, which assumes independence between the first non-fatal event process and the potential death processes, conditional on per-subject random frailty. Simulation studies are conducted to verify the reliability of our estimators. We illustrate the method using the Carvedilol Prospective Randomized Cumulative Survival trial, which involves heart-failure events.
\end{abstract}


\section{Introduction}
In clinical trials involving both mortality and morbidity, the treatment effect on the observed risk of the first non-fatal event is commonly evaluated using the cause-specific hazard ratio. This ratio is estimated using the Cox proportional hazards model that treats death as non-informative censoring. Since the death process and the first non-fatal event process within the same subject are correlated---and because death prevents the occurrence of a non-fatal event---the active treatment may influence the cause-specific hazard ratio directly by affecting the underlying non-fatal event process, indirectly through the death process, or both. This correlation complicates and introduces uncertainty in the interpretation of the cause-specific hazard ratio. Therefore, except in cases where it is confidently known that the active treatment does not have differential effects on mortality, the ``direct effect'' on the underlying first non-fatal event process that is not mediated by competing risks of death---a modification of the underlying mechanism that produces the first non-fatal event---is desired \citep{young2020causal}. A formal mathematical definition of direct effect will be provided in the following section.

\citet{frangakis2002principal} proposed the principal stratification framework to account for post-randomization outcomes such as death. We focus on the principal stratum estimand for addressing the competing risk of death in a randomized clinical trial. Principal stratification is a partition of the population into subpopulations (strata) based on joint values of the competing outcomes under all treatment conditions. In our context, the principal stratum estimand targets subjects who would survive to a specific time point $t$, irrespective of their treatment assignment, within the intended study population. This subpopulation constitutes the principal stratum of interest, termed ``Always Survivors'' defined at $t$, and is denoted as $\mathcal{A}(t)$. The treatment effect within $\mathcal{A}(t)$ is often referred to as the Survivor Average Causal Effect (SACE) \citep{rubin2006causal}.

Typically, $\mathcal{A}(t)$ is defined using a limited number of time points, and the SACE is presented across $\mathcal{A}(t)$ defined at varying $t$ \citep{lyu2023bayesian, comment2019survivor}. This mirrors milestone survival analysis, which compares survival probabilities at pre-specified time points rather than considering the entire curve \citep{chen2015milestone}. While milestone analysis is often advocated in immunotherapy trials, where non-proportionality is common and significantly reduces the power of the Cox model, simulation studies suggest the Cox model generally has higher power unless there is a significant treatment effect delay \citep{gregson2019nonproportional}. Moreover, presenting snapshot effects at specific time points does not provide decision-makers with insights into the sensitivity of conclusions to the potentially arbitrary selection of $t$. In trials where violation of the proportionality assumption is not obvious, we expect the Cox model, capturing treatment effects over time, to have higher power and be more informative than snapshot effects. To capture treatment effects across $\mathcal{A}(t)$ defined at different $t$, we extend the hazard function and define the principal stratum hazard as the instantaneous probability of experiencing the non-fatal event given that the subject has not experienced the non-fatal event and death and would survive to $t$ under the counterfactual arm. Additionally, we propose the Proportional Principal Stratum Hazards (PPSH) model, which maintains the structure of the Cox model but replaces the hazard function with the principal stratum hazard function. The PPSH model assumes and estimates a constant hazard ratio shared across $\mathcal{A}(t)$ defined at different $t$. If the PPSH model is correctly specified, the principal stratum hazard ratio reflects the direct effect on the underlying first non-fatal event process, conditional on being in $\mathcal{A}(t)$. We refer to this as the ``conditional direct effect,'' as will be discussed in more detail later.

Unfortunately, we cannot directly observe $\mathcal{A}(t)$ due to its counterfactual nature: if a subject survived to time $t$ under one arm, we cannot determine if they would have survived to time $t$ under the other arm. However, correctly estimating the probability of each subject belonging to $\mathcal{A}(t)$ (referred to as principal stratum probability) enables us to estimate the PSHR. Estimating principal stratum probability requires further assumptions. One commonly used assumption is the monotonicity assumption, which, in our context, implies that for every subject, their potential time to death in the active treatment arm is no shorter than in the placebo arm \citep{angrist1996identification}. While the monotonicity assumption may be scientifically plausible in certain situations, it is important to note that an active treatment delaying mortality for the population does not necessarily mean it delays mortality for every subject. It remains uncertain whether the monotonicity assumption is plausible in the presence of death. Moreover, the monotonicity assumption alone is not sufficient to estimate principal stratum probabilities. Additional assumptions are typically required \citep{ding2017principal,  zehavi2023matching, isenberg2024weighting}.

The challenge in estimating principal stratum probabilities is that the potential death time from the counterfactual arm and the first non-fatal event time cannot be jointly observed, preventing direct identification of the joint survival function. To address this issue, the shared frailty model offers a potential solution. In this model, each subject is assigned a random effect, referred to as frailty. The potential death time from the counterfactual arm and the first non-fatal event time are assumed to be independent, conditional on the frailty. Similar assumptions have been adopted in previous studies to factorize the joint density of potential outcomes \citep{lyu2023bayesian, comment2019survivor}.

The paper is structured as follows: Section \ref{c5:sec:estimand} introduces the principal stratum estimand within our proposed causal estimand framework. Section \ref{c5:sec:estimation} discusses the PPSH model and the identification of principal stratum membership. Section \ref{c5:sec:simulation} presents a simulation study to verify our proposed method. Section \ref{c5:sec:example} demonstrates the application of our method in a real cardiovascular trial. Finally, we conclude with a discussion in section \ref{c5:sec:discussion}.

\section{Estimand}\label{c5:sec:estimand}
We focus on clinical trials where patients are randomly assigned to an active treatment or placebo. In this chapter, with no loss of generality and to avoid extra notation, we will subsequently assume that we are already within cells defined by a specific value of covariates. 

Assume there are $n$ subjects $(i=1, \dots, n)$ in the study. Each subject $i$ is assigned to either the active treatment arm ($Z_i=1$) or the placebo arm ($Z_i=0$). Following causal inference conventions, we impose the Stable Unit Treatment Value Assumption (SUTVA) \citep{rubin1986comment}. SUTVA posits that there is only one version of the active treatment; otherwise, we would need to define potential outcomes corresponding to each version. Additionally, SUTVA assumes no interference between subjects, implying that each potential outcome for the $i$th subject is determined solely by the treatment received by that subject and does not vary with treatments assigned to others. Let $Y_{i}^{z}$ and $T_{i}^{z}$ be independent and identically distributed (IID) realizations of $Y^{z}$ and $T^{z}$, representing the potential time to death and potential time to the first non-fatal event under treatment $z$, respectively. Since death precludes the occurrence of a non-fatal event, $T_{i}^{z}$ is subject to truncation of $Y_{i}^{z}$ \citep{zhang2003estimation}. Let $Y_i$ and $T_i$ be the IID realizations of $Y$ and $T$, representing the observed outcomes.

\begin{assumption} 
\textit{Under SUTVA, the observable variables $Y_{i}$ and $T_{i}$ satisfy $Y_{i}=Y_{i}^{z}$ and $T_{i}=T_{i}^{z}$ if $Z_i=z$.}
\end{assumption}

\subsection{Causal Estimand}\label{causalest}
\cite{hernan2010causal} define a causal effect as a contrast of any functional of the distributions of counterfactual outcomes under different actions or treatment values. This definition can be expressed mathematically as follows. A parameter $\beta$ is a causal estimand if there exists a monotone function $h(\beta)$ and functional $G(F)$ such that:
\begin{eqnarray}
    h(\beta)=G(F_1)-G(F_0) \label{causaldef}
\end{eqnarray}
where $F_{1}$ and $F_{0}$ represent the cumulative distribution functions (CDFs) of counterfactual outcomes of interest, possibly as functions of some baseline covariates. This definition requires that the counterfactual outcome of interests must be totally ordered and defined for all subjects. Different choices of $G(\cdot)$ result in different causal estimands.

Let $F_{z}(t)=P(T^{z} \leq t)$ represent the CDF of the potential first non-fatal event time in the absence of death. $F_{z}(t)$ can be functions of baseline covariates. According to the definition (\ref{causaldef}), $G\{F_1(t)\}-G\{F_0(t)\}$ is a causal estimand for treatment effects on the first non-fatal event in the absence of death and for direct effects on the underlying first non-fatal event process in the presence of death.

In the absence of death, the hazard ratio in the Cox model that treats the first non-fatal event $T$ as the outcome is a causal estimand in a randomized trial, provided that the Cox model is correctly specified. Due to randomization, the pair $(T^1, T^0)$ is independent of the treatment assignment $Z$, then in the Cox model, we have $F_{z}(t)=F(t \mid Z)=1-\hbox{exp}\{ -\hbox{exp}(\beta Z) \Lambda_{0}(t) \}$ where $\Lambda_{0}(t)$ is the baseline cumulative hazard function. Let $G\{F(t\mid Z)\}=\hbox{log}[ -\hbox{log}\{ 1-F(t\mid Z) \} ]=\hbox{log}\Lambda_{0}(t) + \beta Z$. Then $G\{F_1(t)\}-G\{F_0(t)\}=G\{F(t \mid Z=1) \} - G\{F(t \mid Z=0) \}=\beta$. This may not hold in an observational study, where a patient's prognosis may influence the chosen treatment option. If $F_{z}(t)$ depends on baseline covariates and proportional hazards hold, then $\beta$ represents the conditional hazard ratio, which also has a causal interpretation.

\subsection{Cause-Specific Hazard Ratio (CSHR)}
In the presence of death, $T_{i}^{z}$ is subject to truncation of $Y_{i}^{z}$. A subject is considered at risk for the first non-fatal event at time $t$ if they have not experienced the non-fatal event $(T\geq t)$ and have not died $(Y > t)$. Accordingly, the hazard can be expressed as $\lim_{\Delta t \to 0} P(t \leq T < t+ \Delta t \mid T \geq t, Y > t)/\Delta t$, termed as the cause-specific hazard. The CSHR can then be represented as: 
$$ \lim_{\Delta t \to 0} \frac{P(t \leq T^1 < t+\Delta t \mid T^1 \geq t, Y^{1} > t)}{P(t \leq T^0 < t+\Delta t \mid T^0 \geq t, Y^{0} > t)}.$$ 
In a randomized trial, this ratio can be estimated by:
$$\lim_{\Delta t \to 0} \frac{P(t \leq T < t+\Delta t \mid T \geq t, Y > t, Z=1)}{P(t \leq T < t+\Delta t \mid T \geq t, Y > t, Z=0)}.$$ 

Let $F^{*}_{z}(t)=P(T^z \leq t \mid Y^z >t)$ represent the CDF of the potential first non-fatal event time in the presence of death. By defining $G\{F^{*}_{z}(t)\}=\hbox{log}[ -\hbox{log}\{ 1-F^{*}_{z}(t) \} ]$, the contrast $G\{F^{*}_{1}(t)\}-G\{F^{*}_{0}(t)\}$ corresponds to the CSHR in the Cox model. Due to correlation between $Y^z$ and $T^z$, $F^{*}_{z}(t) \neq F_{z}(t)$. As a result, with the same transformation $G(\cdot)$, $G\{F^{*}_{1}(t)\}-G\{F^{*}_{0}(t)\} \neq G\{F_{1}(t)\}-G\{F_{0}(t)\}$. Thus, the CSHR in the presence of death does not correspond to direct effects on the underlying first non-fatal event process, unless the population is homogeneous, implying $Y^z$ and $T^z$ are independent and $F^{*}_{z}(t) = F_{z}(t)$.

If the active treatment exhibits no differential effects on survival (i.e., $Y^{1}=Y^{0}$), then 
\begin{eqnarray*}
    &&\lim_{\Delta t \to 0} \frac{P(t \leq T^1 < t+\Delta t \mid T^1 \geq t, Y^{1} > t)}{P(t \leq T^0 < t+\Delta t \mid T^0 \geq t, Y^{0} > t)}\\
    &&=\lim_{\Delta t \to 0} \frac{P(t \leq T^1 < t+\Delta t \mid T^1 \geq t, Y^{1} > t, Y^{0} > t)}{P(t \leq T^0 < t+\Delta t \mid T^0 \geq t, Y^{1} > t, Y^{0} > t)}.
\end{eqnarray*}
Since the potential death times $(Y^{1}, Y^{0})$ are unaffected by the actual treatment assignment, conditioning on $Y^{1} > t$ and $Y^{0} > t$ is akin to conditioning on a baseline covariate. Thus, the CSHR reflects the direct effect conditional on being in $\mathcal{A}(t)$ (the conditional direct effect). However, if the active treatment has differential effects on survival, this hazard ratio does not correspond to the conditional direct effect and lacks a causal interpretation. Nonetheless, this observation motivates us to define the PSHR, as outlined below.

\subsection{Principal Stratum Hazard Ratio (PSHR)}
Under SUTVA, at each time $t$, subjects can be classified into four strata according to a pair of counterfactual death time $(Y_{i}^{1}, Y_{i}^{0})$:
\begin{itemize}
    \item Always Survivors: $\{i \mid Y_{i}^{1} > t, Y_{i}^{0} > t \}$, subjects who would survive to $t$ regardless of treatment assignment.
    \item Never Survivors: $\{i \mid Y_{i}^{1} \leq t, Y_{i}^{0} \leq t \}$, subjects who would not survive to $t$ regardless of treatment assignment.
    \item Active Survivors: $\{i \mid Y_{i}^{1} > t, Y_{i}^{0} \leq t \}$, subjects who would survive to $t$ under the active treatment arm, but would not survive to $t$ under the placebo arm. 
    \item Placebo Survivors: $\{i \mid Y_{i}^{1} \leq t, Y_{i}^{0} > t \}$, subjects who would survive to $t$ under the placebo arm, but would not survive to $t$ under the active treatment arm. 
\end{itemize}

Among these four strata at $t$, our primary focus is on the ``Always Survivors'' stratum $\mathcal{A}(t)$. Because subjects in $\mathcal{A}(t)$ can survive to $t$ regardless of the treatment, there is no competing risk of death. We define the principal stratum hazard within $\mathcal{A}(t)$, denoted as $\lambda^{P}(t)$, as the instantaneous probability of experiencing the non-fatal event, given that the subject has not experienced either the non-fatal event or death, and would survive to time $t$ under the counterfactual arm. Mathematically, this is represented as:
\begin{eqnarray*}
    \lambda^{P}(t) = \lim_{\Delta t \to 0} \frac{P(t \leq T < t+\Delta t \mid T\geq t, Y^{1} > t, Y^{0} > t)}{\Delta t}. 
\end{eqnarray*}

Note that if there is no mortality in the trial, i.e., $P(Y^{1} > t)=P(Y^{0} > t)=1$, then $\lambda^{P}(t)$ becomes the well-known hazard function $\lim_{\Delta t \to 0} P(t \leq T < t+\Delta t \mid T\geq t)/\Delta t$.

The PSHR can be expressed as
$$ \lim_{\Delta t \to 0} \frac{P(t \leq T^1 < t+ \Delta t \mid T^1 \geq t, Y^{1} > t, Y^{0} > t)}{P(t \leq T^0 < t+\Delta t \mid T^0 \geq t, Y^{1} > t, Y^{0} > t)}.$$

In the absence of death, when $P(Y^{1} > t)=P(Y^{0} > t)=1$, the PSHR simplifies to the hazard ratio. In the presence of death, $(Y^{1} > t, Y^{0} > t)$ defines the hazard ratio within $\mathcal{A}(t)$, where there is no competing risk of death. The principal stratum membership at $t$ is determined based on counterfactual outcomes $Y^1$ and $Y^0$, which are not affected by treatment assignment and therefore can be regarded as a baseline covariate. Thus, the PSHR in $\mathcal{A}(t)$ at time $t$, where there is no competing risk of death, can be considered as the conditional hazard ratio with principal stratum membership at time $t$ as the baseline covariate, and therefore, reflects the conditional direct effect on the underlying first non-fatal event process. 

The PSHR shares a similar idea with the concept of the causal hazard ratio introduced by \cite{martinussen2020subtleties} which is defined within the subpopulation of those who will potentially survive until a specific time, regardless of the treatment received. While \cite{martinussen2020subtleties} focused on the treatment effect on mortality, our focus is on the direct effect on the underlying first non-fatal event process. Both definitions are related to the SACE \citep{rubin2006causal}. Further discussion on SACE can be found in section \ref{c5:sec:discussion}.

Following the structure of the Cox model, the PPSH model specifies that
\begin{eqnarray}
    \lambda^{P}(t \mid Z) = \hbox{exp}(\beta Z)  \lambda^{P}_{0}(t) \label{psphform}
\end{eqnarray}
where $\lambda^{P}_{0}(t)$ is the (marginal) baseline principal stratum hazard function, and 
$$\hbox{exp}(\beta)=\frac{\lambda^{P}(t \mid Z=1)}{\lambda^{P}(t \mid Z=0)}= \lim_{\Delta t \to 0}
    \frac{P(t \leq T < t+\Delta t \mid T\geq t, Y^{1} > t, Y^{0} > t, Z=1)}{P(t \leq T < t+\Delta t \mid T\geq t, Y^{1} > t, Y^{0} > t, Z=0)}$$
which provides an estimate of the PSHR in a randomized trial. If the PPSH model is correctly specified, the PSHR, which reflects the conditional direct effect on the underlying first non-fatal event process, has a causal interpretation.

\section{Estimation of the PSHR}\label{c5:sec:estimation}
\subsection{Principal Stratum Membership}\label{membershipest}
The actual identification of $\mathcal{A}(t)$ is generally not possible. However, with assumptions and modeling, we can estimate the probability of each subject $i$ being in $\mathcal{A}(t)$ (principal stratum probability). Here, we introduce one potential approach to estimate this probability.

Let $C_{i}$ be the IID realizations of the censoring time $C$. We assume censoring time is independent of potential outcomes:
\begin{assumption}
    $(T^1, T^0, Y^1, Y^0) \indep C$
\end{assumption}

Let $\mathcal{S}$ represent the set of the first non-fatal event times from all subjects. Suppose there are $m$ event times within the set $\mathcal{S}$. We denote these unique event times as $0 < t_1 < t_2 < \dots < t_m$. At each $t_j \in \mathcal{S}$, we define $\mathcal{R}_{j}$ as the at-risk set at $t_j$, which comprises subjects who remain in the study $(Y_i > t_j, C_i > t_j)$ and have not yet experienced a non-fatal event by time $t_j$ $(T_i \geq t_j)$. Each subject $i$ belonging to $\mathcal{R}_{j}$, whose last known follow-up time is denoted by $D_i=\hbox{min}(Y_i, C_i)$ with realization $d_i$, where $d_i > t_j$, falls into one of the following two cases:

\begin{enumerate}
    \item $T_i=t_j$: The first non-fatal event occurred at $t_j$.
    \item $T_i>t_j$: The first non-fatal event has not yet occurred by $t_j$.
\end{enumerate}

Subject $i$ belonging to $\mathcal{R}_{j}$ is associated with a principal stratum probability of being in $\mathcal{A}(t_j)$, denoted by $p_{ij}$. Suppose, without loss of generality, that a subject $i$ is assigned to arm $z$. 

If the subject $i$ falls into case 1, the principal stratum probability at $t_j$ is:
\begin{eqnarray}
    p_{ij} =  P(Y^{1-z}_{i}>t_j \mid T^{z}_i=t_j) = \frac{P(Y^{1-z}_{i}>t_j,  T^{z}_i=t_j )}{P(T^{z}_i=t_j)}. \label{eqnstep1}
\end{eqnarray}

As previously mentioned, the joint distribution of $Y^{1-z}$ and $T^{z}$ cannot be directly identified because these potential outcomes are correlated but never jointly observed. The correlation arises because both $Y^{1-z}$ and $T^{z}$ are outcomes from the same subjects. The correlation can be modeled by a subject-level random effect. Specifically, we assume the correlation between $Y^{1-z}$ and $T^{z}$ is captured by the variability of this random effect (e.g., its variance), allowing us to model them as conditionally independent given the random effect. In time-to-event analysis, such random effects are known as frailties, and the resulting models are referred to as shared frailty models \citep{balan2020tutorial}. To factorize the joint distribution appearing in the numerator of equation (\ref{eqnstep1}), we adopt the following conditional independence assumption:
\begin{assumption}
    Conditional on the frailty $\theta$, the potential death time from the counterfactual arm is independent of the potential first non-fatal event time: $Y^{1-z} \indep T^{z} \mid \theta$, for $z=0, 1$. \label{assumption:condind1}
\end{assumption}

While this conditional independence assumption differs slightly from those in \cite{lyu2023bayesian} and \cite{comment2019survivor}, it shares a similar motivation. In their frameworks, the assumption is that the potential outcomes under different treatment arms are independent conditional on $\theta$, i.e., $(Y^{1-z}, T^{1-z}) \indep (Y^z, T^z) \mid \theta$, which appears to be a somewhat stronger assumption than ours.

Using the imposed conditional independence assumption, we can factorize equation (\ref{eqnstep1}), as follows:
\begin{eqnarray*}
    p_{ij} = \frac{\int P(Y^{1-z}_{i}>t_j, T^{z}_i=t_j \mid \theta) p(\theta) d\theta}{\int P(T^{z}_i=t_j \mid \theta) p(\theta) d\theta} \nonumber = \frac{\int P(Y^{1-z}_{i}>t_j \mid \theta) P(T^{z}_i=t_j \mid \theta) p(\theta) d\theta}{\int P(T^{z}_i=t_j \mid \theta)p(\theta) d\theta} 
\end{eqnarray*}
where $\theta$ is the frailty and $p(\theta)$ is the density function of $\theta$. 

Let $\eta^{z}_{Y}(t)$ and $\eta^{z}_{T}(t)$ denote the conditional cumulative hazard of $Y^{z}$ and $T^{z}$ given $\theta=1$, then
\begin{eqnarray*}
    p_{ij}
    &=& \frac{\int \hbox{exp}\{-\theta \eta^{1-z}_{Y}(t_j)\} \theta \hbox{exp}\{-\theta \eta^{z}_{T}(t_j)\} p(\theta) d\theta}{\int \theta \hbox{exp}\{-\theta \eta^{z}_{T}(t_j)\} p(\theta) d\theta} .
\end{eqnarray*}

For an explicit expression for the principal stratum probability, with practical considerations, we assume $\theta \sim \Gamma(\gamma, \gamma)$ where the mean is 1, and the variance is $1/\gamma$. Then:
\begin{eqnarray*}
    p_{ij}
    &=& \frac{\int \hbox{exp}\{-\theta \eta^{1-z}_{Y}(t_j)\} \theta \hbox{exp}\{-\theta \eta^{z}_{T}(t_j)\} \frac{\gamma^{\gamma}}{\Gamma(\gamma)} \theta^{\gamma -1} \hbox{exp}(-\gamma \theta) d\theta}{\int \theta \hbox{exp}\{-\theta \eta^{z}_{T}(t_j)\} \frac{\gamma^{\gamma}}{\Gamma(\gamma)} \theta^{\gamma -1} \hbox{exp}(-\gamma \theta) d\theta} \\
    &=& \frac{ \int \frac{\gamma^{\gamma}}{\Gamma(\gamma)}  \theta^{\gamma} \hbox{exp}[-\theta \{ \gamma + \eta^{1-z}_{Y}(t_j) + \eta^{z}_{T}(t_j) \} ] d\theta }{ \int \frac{\gamma^{\gamma}}{\Gamma(\gamma)}  \theta^{\gamma} \hbox{exp}[-\theta \{ \gamma + \eta^{z}_{T}(t_j) \} ] d\theta } \\
    &=& \frac{ \left\{ \frac{\gamma}{\gamma + \eta^{1-z}_{Y}(t_j) + \eta^{z}_{T}(t_j)} \right\}^{\gamma+1} }{\left\{\frac{\gamma}{\gamma + \eta^{z}_{T}(t_j)} \right\}^{\gamma+1} } = \left\{  \frac{\gamma +  \eta^{z}_{T}(t_j)}{\gamma + \eta^{1-z}_{Y}(t_j) + \eta^{z}_{T}(t_j)}\right\}^{\gamma +1}.
\end{eqnarray*}

Similarly, if the subject $i$ falls into case 2, the principal stratum probability at $t_j$ is:
$$p_{ij}=P(Y^{1-z}_{i}>t_j \mid T^z_i>t_j) = \left\{  \frac{\gamma + \eta^{z}_{T}(t_j)}{\gamma + \eta^{1-z}_{Y}(t_j) + \eta^{z}_{T}(t_j)}\right\}^{\gamma}.$$

Assuming a distribution for $\theta$ may allow us to derive conditional cumulative hazards $\eta^{z}_{Y}(t)$ and $\eta^{z}_{T}(t)$ from marginal functions $P(Y^z > t)$ and $P(T^z > t \mid Y^z > t)$. As above, assuming $\theta \sim \Gamma(\gamma, \gamma)$ and let $S^{z}_{Y}(t)=P(Y^z > t)$, we have:
\begin{eqnarray*}
    S^{z}_{Y}(t)=\int \hbox{exp}\{-\theta \eta^{z}_{Y}(t)\}p(\theta)d\theta &=& \int \hbox{exp}\{-\theta \eta^{z}_{Y}(t)\}\frac{\gamma^{\gamma}}{\Gamma(\gamma)} \theta^{\gamma -1} \hbox{exp}(-\gamma \theta) d\theta \\
    &=& \left\{ \frac{\gamma}{\gamma+\eta^{z}_{Y}(t)} \right\}^{\gamma}.
\end{eqnarray*}

Then, the conditional cumulative hazard of $Y^{z}$ given $\theta$ = 1 is
\begin{eqnarray}
\eta^{z}_{Y}(t) = \gamma \{S^{z}_{Y}(t)^{-1/\gamma} -1\} .\label{latentY} 
\end{eqnarray}

The conditional cumulative hazard of $T^{z}$ given $\theta$ = 1 is derived similarly, but we must account for the competing risk of death. Let $S^{z}_{T}(t \mid Y^{z} > t)=P(T^{z} > t \mid Y^{z} > t)$, we have 
\begin{eqnarray*}
    S^{z}_{T}(t \mid Y^{z} > t)=\int P(T^z > t \mid \theta, Y^z > t) p(\theta \mid Y^z > t)d\theta .
\end{eqnarray*}

If we assume $P(T^z > t \mid \theta, Y^z > t)=P(T^z > t \mid \theta)$, then the expression simplifies to
\begin{eqnarray*}
    S^{z}_{T}(t \mid Y^{z} > t)=\int P(T^z > t \mid \theta) p(\theta \mid Y^z > t)d\theta = \int \hbox{exp}\{-\theta \eta^{z}_{T}(t)\}p(\theta \mid Y^z > t)d\theta = \left\{ \frac{\gamma + \eta^{z}_{Y}(t)}{\gamma+\eta^{z}_{Y}(t) + \eta^{z}_{T}(t)} \right\}^{\gamma}
\end{eqnarray*}
where $p(\theta \mid Y^z > t)$ is available in closed form only if $\theta$ follows a gamma distribution \citep{balan2020tutorial}.

Using this result, the conditional cumulative hazard of $T^{z}$ given $\theta$ = 1 is
\begin{eqnarray}
    \eta^{z}_{T}(t) = \{\gamma+\eta^{z}_{Y}(t) \}  \{S^{z}_{T}(t \mid Y^z > t)^{-1/\gamma}-1\} .\label{latentT}
\end{eqnarray}

Therefore, obtaining a closed-form expression for $\eta^{z}_{T}(t)$ requires the assumption $P(T^z > t \mid \theta, Y^z > t)=P(T^z > t \mid \theta)$, which implies that, conditional on $\theta$, the distribution of $Y^z$ is independent of $T^z$. This leads to the following assumption:
\begin{assumption}
    Conditional on the frailty $\theta$, the potential death time is independent of the potential first non-fatal event time: $Y^{z} \indep T^{z} \mid \theta$, for $z=0, 1$. \label{assumption:condind2}
\end{assumption}
This assumption has also been used in \cite{lyu2023bayesian} and \cite{comment2019survivor} to factorize their joint likelihood functions.

Assumptions~\ref{assumption:condind1} and \ref{assumption:condind2} together imply that the potential first non-fatal event time is conditionally independent of the potential death times under both treatment arms, given that the frailty appropriately accounts for the correlation. Notably, we do not impose any conditional independence assumption between the two potential death times, $Y^z$ and $Y^{1-z}$, across arms.

The marginal functions $S^{z}_{Y}(t)$ and $S^{z}_{T}(t \mid Y^{z} > t)$ are not directly observable but, in a randomized trial, they can be estimated by $S_{Y}(t \mid Z=z)$ and $S_{T}(t \mid Y > t, Z=z)$, respectively. This is because randomization guarantees independence between treatment assignment $Z$ and potential outcomes. Therefore, we have:
$$S_{Y}(t \mid Z=z) = P(Y > t \mid Z=z) = P(Y^{z} > t \mid Z=z) = P(Y^{z} > t)=S^{z}_{Y}(t).$$
Similarly, we have
$S_{T}(t \mid Y > t, Z=z)=S^{z}_{T}(t \mid Y^{z} > t)$.

To estimate $S_{Y}(t \mid Z=z)$, in practice, we fit a Cox model where we view death $Y$ as the outcome and the treatment assignment $Z$ as the covariate. $S_{T}(t \mid Y > t, Z=z)$ can be estimated nonparametrically as $\sum_{Z_i=z}I(T_i >t)\big/\sum_{Z_i=z}I(Y_i >t)$, where $I(\cdot)$ is the indicator function.

We assume $\theta \sim \Gamma(\gamma, \gamma)$ and use $\theta$ to induce correlation between $T^{z}$ and $Y^{1-z}$. However, $T^{z}$ and $Y^{1-z}$ are never observed jointly, so we need to pre-specify some values of $\gamma$ when calculating the principal stratum probabilities. The estimation of principal stratum hazard ratio, which will be addressed in section \ref{secpcpeffects}, needs estimated principal stratum probabilities as input. Therefore, we can expect the estimated principal stratum hazard ratio to vary with specified $\gamma$. While one might consider estimating $\gamma$---since $\theta$ also induces correlation between the observed non-fatal event and death times---the goal of the PPSH model is not to produce a single point estimate, but rather to complement the cause-specific hazard ratio with a range of plausible principal stratum hazard ratios based on pre-specified values of $\gamma$. Moreover, estimating $\gamma$ faces numerous technical issues, as detailed in section A of the Supplementary Materials.

In summary, we present one potential approach to estimate principal stratum probabilities under the conditional independence assumption and the independent censoring assumption in a randomized trial. The estimation consists of three steps:
\begin{enumerate}
    \item Fit the Cox model for death time $Y$ to obtain the estimated $\hat S_{Y}(t \mid Z=z)$. Estimate $S_{T}(t \mid Y > t, Z=z)$ nonparametrically as $\hat S_{T}(t \mid Y > t, Z=z)=\sum_{Z_i=z}I(T_i >t)\big/\sum_{Z_i=z}I(Y_i >t)$. $\hat S_{Y}(t \mid Z=z)$ and $\hat S_{T}(t \mid Y > t, Z=z)$ serve as the estimates of the marginal functions $S^{z}_{Y}(t)$ and $S^{z}_{T}(t \mid Y^{z} > t)$, respectively.    
    \item Estimate the conditional cumulative hazard functions $\eta^{z}_{Y}(t)$ and $\eta^{z}_{T}(t)$ from the estimated marginal functions $\hat S^{z}_{Y}(t)$ and $\hat S^{z}_{T}(t \mid Y^{z} > t)$ using equations (\ref{latentY}) and (\ref{latentT}).
    \item Collect the first non-fatal event times from all subjects into set $\mathcal{S}$. At each $t_j \in \mathcal{S}$, we calculate the principal stratum probability for each subject in $\mathcal{R}_j=\{i \mid Y_i > t_j, C_i > t_j, T_i \geq t_j \}$, using the proposed formula with $\hat \eta^{z}_{Y}(t)$ and $\hat \eta^{z}_{T}(t)$ from step 2 and the pre-specified $\gamma$. The formula to use depends on which case the subject belongs to.
\end{enumerate}

\subsection{Proportional Principal Stratum Hazards (PPSH) Model}\label{secpcpeffects}
The likelihood function of the PPSH model resembles that of the Cox model but incorporates certain modifications. Following the notation in section \ref{membershipest}, suppose there are $m$ unique event times $0 < t_1 < t_2 < \dots < t_m$ within the set $\mathcal{S}$. We first consider the scenario where there are no tied events. That is, only one subject had a non-fatal event at $t_j$, and we index this subject by $(j)$. Additionally, we assume that this subject belongs to the treatment arm $z$. Let $\mathcal{V}=\{ t_j \in S: Y^{1-z}_{(j)} > t_j \}$. In other words, for each $t_j$, where $j=1,\dots,m$, if the subject $(j)$ who experienced a non-fatal event at $t_j$ belongs to $\mathcal{A}(t_j)$, $t_j$ is a member of $\mathcal{V}$; otherwise, it is not. Since the principal stratum hazard is defined within $\mathcal{A}(t_j)$, the partial likelihood (PL) is constructed within $\mathcal{V}$. Suppose the value of $p_{ij}$ is known, we have:
\begin{eqnarray*}
    PL(\beta) &=& \prod_{j: t_j \in \mathcal{V}}  \frac{p_{(j)j}  \lambda^{P}\{t_j \mid Z_{(j)}\} \Delta t_{j}}{\sum_{i \in \mathcal{R}_j} p_{ij}  \lambda^{P}(t_j \mid Z_{i}) \Delta t_{j}} = \prod_{j: t_j \in \mathcal{V}} \frac{p_{(j)j}  \hbox{exp}\{\beta Z_{(j)}\}}{\sum_{i \in \mathcal{R}_j} p_{ij} \hbox{exp}(\beta Z_{i})}.
\end{eqnarray*}
The second equation holds due to the structure of the PPSH model (\ref{psphform}), where the baseline $\lambda^{P}_0(t)$ is dispensed. Then the log partial likelihood is 
    $$pl(\beta)= \hbox{log}PL(\beta) = \sum_{j: t_j \in \mathcal{V}} \left\{ \hbox{log}p_{(j)j} + \beta Z_{(j)} - \hbox{log}\sum_{i \in \mathcal{R}_j} p_{ij} \hbox{exp}(\beta Z_{i}) \right\}.$$

In the case of tied events (more than one subject had a non-fatal event at $t_j$), Breslow suggests using the same expression for $pl(\beta)$ \citep{breslow1974covariance}.

Let 
$$W_j=Z_{(j)} - \frac{\sum_{i \in \mathcal{R}_j} p_{ij}  Z_{i} \hbox{exp}(\beta Z_{i})}{\sum_{i \in \mathcal{R}_j} p_{ij} \hbox{exp}(\beta Z_{i})}.$$ 

The partial likelihood score function is 
$$U(\beta)=\frac{\partial pl(\beta)}{\partial \beta} =\sum_{j: t_j \in \mathcal{V}} W_j = \sum_{j: t_j \in \mathcal{S}} I\{Y^{1-z}_{(j)} > t_j\}  W_j$$
where $I\{Y^{1-z}_{(j)} > t_j\}$ is the indicator that the subject $(j)$ belongs to $\mathcal{A}(t_j)$.

We can estimate $\beta$ by solving $U(\beta)=0$. However, $\mathcal{V}$ is non-identifiable because we cannot observe $I\{Y^{1-z}_{(j)} > t_j\}$. To address this non-identifiability issue, we introduce the modified score function: 
\begin{eqnarray*}
    U^{*}(\beta)=\sum_{j: t_j \in \mathcal{S}} p_{(j)j}  W_j \label{modscore}.
\end{eqnarray*}

We can demonstrate that, for $j: t_j \in \mathcal{S}$,
\begin{eqnarray*}
    \hbox{E}\left[ I\{Y^{1-z}_{(j)} > t_j\}  W_j \right] &=& \hbox{E} \left( \hbox{E} \left[ I\{Y^{1-z}_{(j)} > t_j\}  W_j \mid \theta_{j} \right] \right) \\
    &=&  \hbox{E} \left( \hbox{E} \left[ I\{Y^{1-z}_{(j)} > t_j\} \mid \theta_{j} \right]  \hbox{E} \left( W_j \mid \theta_{j} \right) \right) \\
    &=&  \hbox{E} \left\{ p_{(j)j}  \hbox{E} \left( W_j \mid \theta_{j} \right) \right\} =  p_{(j)j}   \hbox{E} \left\{  \hbox{E} \left( W_j \mid \theta_{j} \right) \right\} \\
    &=& p_{(j)j}    \hbox{E} \left( W_j \right) = \hbox{E} \left\{ p_{(j)j}  W_j \right\}
\end{eqnarray*}
where the second equation holds because of the principal ignoribility assumption.

Therefore, instead of solving $U(\beta) = 0$, we can estimate $\beta$ by solving $U^{*}(\beta)=0$. The standard Newton-Raphson algorithm can be employed to find a solution to $U^{*}(\beta)=0$. Specifically, starting with an initial guess $\hat \beta^{(0)}$, the algorithm iteratively computes
$$\hat \beta^{(n+1)} = \hat \beta^{(n)}+\mathcal{L}^{-1}\{\hat \beta^{(n)}\} U^{*}\{\hat \beta^{(n)}\}$$
until convergence, where $\mathcal{L}(\beta)$ is the negative second derivative of log partial likelihood, given by $\mathcal{L}(\beta) = \sum_{j: t_j \in \mathcal{S}} V(t_j;\beta)$ with
\begin{eqnarray*}
  &&V(t_j;\beta)\\
  &&=p_{(j)j}  \frac{ \left\{ \sum_{i \in \mathcal{R}_j} p_{ij}  \hbox{exp}(\beta Z_{i}) \right\} \left\{ \sum_{i \in \mathcal{R}_j} p_{ij}  Z_{i}^2  \hbox{exp}(\beta Z_{i}) \right\} - \left\{ \sum_{i \in \mathcal{R}_j} p_{ij}  Z_{i}  \hbox{exp}(\beta Z_{i}) \right\}^2 }{\left\{ \sum_{i \in \mathcal{R}_j} p_{ij}  \hbox{exp}(\beta Z_{i}) \right\}^2}.  
\end{eqnarray*}

Given that the objective function exhibits strict concavity, the algorithm is likely to maintain numerical stability and converge quickly toward $\hat \beta$. Convergence problems are very rare using the default initial value of $\hat \beta^{(0)}=0$ \citep{therneau2000cox}.

In summary, the estimation of the PSHR consists of two stages. In stage 1, at each non-fatal event time, we estimate principal stratum probabilities for each at-risk subject. In stage 2, a PPSH model, essentially a weighted Cox model with principal stratum probabilities as weights, is fitted. Since principal stratum probabilities are estimated rather than observed, we replace them with the estimates from stage 1.

To address the variability introduced by estimating principal stratum probabilities, we construct bootstrap confidence intervals (CI) using the following steps:
\begin{enumerate}
    \item Resample subjects with replacement to generate a random bootstrap sample of the same size as the original dataset. Resampling with replacement will include duplicate records from the same subject, and we treat each sampled record as coming from a distinct subject for bootstrap purposes. Accordingly, we assign a unique new subject ID to each record in the bootstrap sample, regardless of whether it originates from the same subject in the original data.
    \item Estimate the principal stratum hazard ratio from the bootstrap sample. This includes estimating principal stratum probabilities and fitting the PPSH model using those estimates.
    \item Repeat steps 1 and 2 for a total of $B$ bootstrap replicates. Let $\{\hat \beta_{(1)}, \hat \beta_{(2)}, \dots, \hat \beta_{(B)}\}$ denote the bootstrap estimates of the principal stratum hazard ratio. Denote $\beta^{*}_{(\alpha/2)}$ and $\beta^{*}_{(1-\alpha/2)}$ as the $100(\alpha/2)\%$ and $100(1-\alpha/2)\%$ of the bootstrap estimates. The resulting $100(1 - \alpha)\%$ bootstrap CI is $(\beta^{*}_{(\alpha/2)}, \beta^{*}_{(1-\alpha/2)})$, using the percentile method \citep{efron1982jackknife}. 
\end{enumerate}

The estimated principal stratum probabilities in section \ref{membershipest} are conditional on specified $\gamma$. As a result, the estimated PSHR is also conditional on $\gamma$. With different specified $\gamma$, we can expect different estimates of $\beta$. Sensitivity analysis is recommended to provide researchers with a range of causal estimates $\hat \beta$ under different values of $\gamma$.

A key assumption of the PPSH model is proportionality. A formal proportionality test is presented in section B of the Supplementary Materials.

\section{Simulations}\label{c5:sec:simulation}
\subsection{Settings}\label{simusetting}
Assuming that $\theta_{i}$ follows a gamma distribution with a mean of one and a variance of $1/\gamma$, we model the potential death time $Y^{z}_{i}$, given $\theta_{i}$, for simulation purposes, as an exponential distribution with a mean of $1/\theta_i \lambda_{z}$. This yields a survival function denoted as $S(t;\theta_i)=\exp(-\theta_{i} \lambda_{z} t)$. The rates for the placebo and active treatment arms are denoted as $\lambda_0$ and $\lambda_1$, respectively. Additionally, we consider the independent loss to follow-up $C$, which follows an exponential distribution with a mean of $1/\lambda_c$. The maximum follow-up time for patients is denoted as $\tau$. The last known follow-up time is then calculated as $D=\min(Y, C, \tau)$.

The model (\ref{psphform}) focuses on the marginal principal stratum hazard function $\lambda^{P}(t \mid Z)$. By defining $\eta^{0}_{T}(t)=\phi t$, and
$$\eta^{1}_{T}(t)=\frac{\phi}{\lambda_{Y}-\phi(r^P-1)} \left[ (1-r^P) \gamma^{1-\frac{r^P\phi}{\phi + \lambda_{Y}}} \{ \gamma + (\phi + \lambda_{Y})t \}^{\frac{r^P\phi}{\phi + \lambda_{Y}}} - \gamma +\gamma r^P + \lambda_{Y}r^Pt \right]$$
where $\lambda_{Y}=\lambda_{0}+\lambda_{1}$, we achieve $\frac{\lambda^{P}(t \mid Z=1)}{\lambda^{P}(t \mid Z=0)}=r^P$, representing a marginal proportional PSHR of $r^P$ for the first non-fatal event. For detailed derivation of $\eta^{1}_{T}(t)$, please refer to the section C of the Supplementary Materials.

To generate non-fatal event time $T^{z}_{i}$ from $\eta^{z}_{T}(t)$, we solve $\theta \eta^{z}_{T}(t)+\log U=0$, where $U \sim \text{Unif}(0,1)$. This approach works because $-\log U \sim \text{Exp}(1)$, and the term $1-\exp\{-\theta \eta^{z}_{T}(t)\}$ represents the CDF of an $\text{Exp}(1)$ distribution. If the generated time $T_i > D_i$, there is no observed non-fatal event during the follow-up for subject $i$.

This simulation algorithm introduces correlations between $(Y^{z}, Y^{1-z}, T^{z})$ through $\theta$. The three processes are then correlated but are mutually independent conditional on $\theta$.

We also generate a separate hypothetical dataset where there is no mortality by setting $\lambda_0=\lambda_1=0$, following the same simulation algorithm as described earlier, with the only exception that each subject's last follow-up time is now calculated as $D=\hbox{min}(C, \tau)$. As discussed, when there is no mortality, the principal stratum hazard function simplifies to the well-known hazard function. Therefore, for this hypothetical dataset, we could expect the Cox model to provide an estimated hazard ratio of $r^P$. 

\subsection{Results}\label{resultssec}
In table \ref{results}, we present a comparison of estimation results between our proposed principal stratum (PS) approach and the cause-specific (CS) approach, with the estimation results from the hypothetical dataset as a reference. Estimation on the hypothetical dataset is conducted by treating the non-fatal event as the outcome and applying the Cox model. To obtain the CS estimate, we also use the Cox model and treat the non-fatal event as the outcome, but relabel death as censoring.

We use common parameter values for this comparison: $\lambda_1=0.2$, $\lambda_c=0.03$, $\tau=2$, $\phi=2$, $r=0.5$. The sample size is 300 with equal allocation. We explore different values of $\lambda_0$ and $\gamma$. With $\lambda_1=0.2$, $\lambda_0=0.4$ indicates remarkable effects on mortality (hazard ratio of 0.5 conditional on $\theta$), while $\lambda_0=0.25$ represents a more realistic yet impressive effect (hazard ratio of 0.8 conditional on $\theta$). $\gamma=5$ denotes a highly homogeneous population, while $\gamma=0.5$ signifies high heterogeneity. For each parameter configuration, we conduct 1000 simulations and provide summary statistics for the datasets used to estimate $\beta$. 

In table \ref{results}, when increasing $\lambda_0$ from 0.25 to 0.4---thereby raising the death rate in the placebo arm---we observe a smaller proportion of subjects experiencing a non-fatal event in that arm, for the same value of $\gamma$ and with $r=0.5$ held constant. This occurs because the non-fatal event process is subject to the competing risk of death. As $\lambda_0$ increases, the time to death in general decreases, leading to more subjects dying before they have a chance to experience a non-fatal event. This demonstrates that even if the treatment effect on the underlying non-fatal event process remains the same, altering the treatment effect on the death process can change the observed risk ratio for the first non-fatal event.

The hypothetical dataset is generated separately using the same parameters except that $\lambda_0=\lambda_1=0$. We define the Monte Carlo bias as the Monte Carlo sample mean of estimates minus the true theoretical value $\hbox{log}(0.5)$. The HR refers to the estimated hazard ratio, computed as the exponential of the Monte Carlo sample mean of estimates. As expected, the Monte Carlo bias from the hypothetical dataset is minimal, and the HR is close to its nominal value.

In the ideal scenario where the true value of $\gamma$ is known, the Monte Carlo bias in our proposed PS approach is also minimal across all parameter configurations, likely due to random variability. In reality, where the true value of $\gamma$ is unknown, misspecifying $\Tilde{\gamma}$ in the PS estimation, leads to bias, and the bias increases as $\Tilde{\gamma}$ deviates further from the true value. 

The bias increases for both PS and CS methods with greater treatment effects on mortality. The relative magnitude of bias between the two approaches is similar across different $\lambda_0$ but varies with different $\gamma$. When the population is highly heterogeneous $(\gamma=0.5)$, bias in CS estimates is pronounced and greater than that from the PS estimates, even when specifying $\Tilde{\gamma}=5$, far from the true value of 0.5. However, bias in CS estimates is less pronounced as $\gamma$ increases. In a highly homogeneous population $(\gamma=5)$, bias in CS estimates becomes minimal and comparable to that from the PS approach.

This phenomenon is understandable considering the origin of bias in the CS approach. Each subject in our simulations has a frailty, with those having higher frailty experiencing mortality and the non-fatal event earlier. The CS approach compares subjects conditional on survival, but since treatment prolongs survival, those in the treatment arm are expected to have higher average frailty and thus a greater risk of the non-fatal event. Consequently, CS estimates tend to be greater than the true value. As treatment effects on mortality become more substantial, the disparity in frailty between the two arms after conditioning on survival becomes more pronounced, leading to increased bias. However, in highly homogeneous populations, the frailty of each subject is less distinguishable, resulting in comparable frailty between the two arms even after conditioning on survival, and thus, comparable risk of the non-fatal event. Therefore, CS estimates are closer to the true value in such scenarios.

In table \ref{results}, with a heterogeneous population $(\gamma=0.5)$, the PSHR approaches the CSHR as $\Tilde{\gamma}$ increases. Setting a large value for $\Tilde{\gamma}$ in the PS approach assumes a highly homogeneous population with similar frailty. In a homogeneous population, even after conditioning on the death time, survivors remain comparable in frailty between both arms. Therefore, for this assumed highly homogeneous population, the PS approach would yield an estimated HR similar to that of the CS approach.

\subsection{Sensitivity Analysis}
When calculating the principal stratum probabilities, we assume the frailty follows a gamma distribution. This choice is common for several reasons. First, the conditional distribution of frailty, given survival, is only available in closed form when frailty follows a gamma distribution \citep{balan2020tutorial}. Second, theoretical results indicate that the gamma distribution is the limiting distribution of the frailty of long-time survivors, irrespective of the frailty distribution at baseline \citep{abbring2007unobserved}. However, estimators of the PPSH model may be biased if the functional form of the frailty distribution is misspecified. Nonetheless, empirical evidence suggests that this bias is generally minimal. Simulation studies indicate that the gamma distribution is robust to such misspecification in terms of bias and efficiency. Specifically, when the true frailty distribution is either inverse Gaussian or positive stable, the estimates from a Cox model with an assumed gamma frailty distribution are minimally affected \citep{gorfine2012conditional, hsu2007robustness, axelrod2023sensitivity}.

In this sensitivity analysis, we generate datasets assuming frailty follows an inverse Gaussian distribution at baseline, with a mean of 1 and a variance of $1/\gamma$. All other parameter settings, simulation algorithms, and estimation procedures, including assuming a gamma distribution for frailty in estimation, remain unchanged.

It is worth noting that, if the frailty used in data generating follows a gamma distribution, the non-fatal event generating algorithm described in section \ref{simusetting} by theory will provide a proportional principal stratum hazard, and when there is no mortality, it should provide a proportional hazard. However, if the frailty used in data generating follows an inverse Gaussian distribution, the data generating algorithm described in section \ref{simusetting} does not, in theory, provide a marginally proportional principal stratum hazard, and there is no explicit formula by which we can generate a marginally proportional principal stratum hazard either. Nonetheless, the inverse Gaussian distribution shares a similar shape with the gamma distribution, so we could expect that even if we generate the datasets using the algorithm in section \ref{simusetting} with inverse Gaussian distributed frailty, the proportionality assumption may not be severely violated. In fact, we use this algorithm to generate datasets from inverse Gaussian distributed frailty in the absence of mortality with other parameter values specified in section \ref{simusetting} and test the proportional hazards assumption with the R function \textit{cox.zph}. At the $\alpha=0.05$ level, there is no violation of the proportional hazards assumption 84\%, 93\%, and 95\% out of 10000 simulations when $\gamma=0.5$, $\gamma=2$, and $\gamma=5$, respectively. As a comparison, if the gamma distributed frailty is used in data generating, 95\% out of 10000 simulations, there is no violation of the proportionality assumption across all $\gamma$, at $\alpha=0.05$ level. Therefore, with inverse Gaussian distributed frailty, although the proportionality assumption is not perfectly protected, we could still expect the estimated principal stratum hazard ratio to provide an informative summary of the treatment effects.  

Table \ref{sensitivityresults} presents results from the sensitivity analysis. Compared to table \ref{results}, when the population is highly heterogeneous $(\gamma=0.5)$, the average proportion of death and subjects with a non-fatal event are lower when the frailty follows a gamma distribution. This discrepancy arises because, with the same mean and variance, the first quartile of the gamma distribution is generally smaller than that of the inverse Gaussian distribution. This suggests that the relatively healthier subjects in the gamma-generated population have a lower probability of death and non-fatal events compared to those in the inverse Gaussian-generated population.

Estimation results from the hypothetical dataset are presented as a reference where $\lambda_0=\lambda_1=0$. We set $r^P=0.5$ in the simulation, but since the frailty now follows an inverse Gaussian distribution, $\hbox{log} r^P$ is no longer the theoretical true value of $\beta$. ``Est'' is the Monte Carlo sample mean of estimates from the hypothetical dataset, serving as the empirical true value of $\beta$. Then the bias is defined as the Monte Carlo sample mean of estimates minus the corresponding empirical true value. In a homogeneous population (large $\gamma$), the difference in frailty distribution has minimal impact. The empirical HR from the hypothetical dataset is close to 0.5, and the bias from both CS and PS approaches is minimal. As population heterogeneity increases, the empirical HR deviates further from 0.5, and the bias of the CS approach becomes pronounced. The bias of the PS is minimal when $\Tilde{\gamma}=\gamma$, and is in general smaller than that from the CS approach when $\Tilde{\gamma}$ is misspecified. Similar to the results in table \ref{results}, when the population is heterogeneous, the PSHR approaches the CSHR as $\Tilde{\gamma}$ increases, for the reason discussed in section \ref{resultssec}.

\section{A Real Example}\label{c5:sec:example}
We illustrate our approach using data from the Carvedilol Prospective Randomized Cumulative Survival (COPERNICUS) trial, a double-blind, placebo-controlled study examining the impact of carvedilol on morbidity and mortality among patients with severe heart failure \citep{packer2001effect}. The trial included 2289 patients in total over a mean period of 10.4 months. Primary outcome analysis showed a 35\% decrease in all-cause mortality with carvedilol. 700 hospitalizations occurred in the carvedilol arm and 848 in the placebo arm. Notably, 38\% had no hospitalizations, with 25\%, 13\%, 10\%, and 14\% experiencing one, two, three, and more than three hospitalizations, respectively. After excluding hospitalizations after the first, 382 hospitalizations remained in the carvedilol arm and 429 in the placebo arm. Compared to trials with mixed-risk populations \citep{teerlink2021cardiac}, the COPERNICUS, focusing on high-risk patients, has a higher proportion experiencing multiple hospitalizations.

Table \ref{copern} presents a comparison of estimation results between the PS and the CS approach. $\Tilde{\gamma}$ represents the value of $\gamma$ employed in the PS approach, indicating the assumed heterogeneity level within the study population. HR of the PS approach is calculated as $\hbox{exp}(\hat \beta)$. 95\% CI of the PS approach is the percentile interval from 1000 bootstraps: $(\hbox{exp}(\beta^{*}_{0.025}), \hbox{exp}(\beta^{*}_{0.975}))$, where $\beta^{*}_{(0.025)}$ and $\beta^{*}_{(0.975)}$ represent the 2.5th and 97.5th percentiles of the bootstrap estimates.

The PSHR is smaller than the CSHR regardless of $\Tilde{\gamma}$. Carvedilol generally extended survival, implying that subjects surviving to time $t$ in the placebo arm were generally healthier at baseline compared to those in the carvedilol arm, and hence expected to be at lower risk of the non-fatal event. The CSHR is thus closer to 1.

In this example, the PSHR is not very sensitive to the choice of $\Tilde{\gamma}$, ranging from 0.781 in an assumed highly heterogeneous population $(\Tilde{\gamma}=0.25)$ to 0.820 in an assumed highly homogeneous population $(\Tilde{\gamma}=10)$. The PSHR approaches the CSHR as $\Tilde{\gamma}$ increases, because the CS approach essentially treats the population as homogeneous, which corresponds to $\Tilde{\gamma} \to \infty$.

The CI from the PS approach narrows with larger $\Tilde{\gamma}$ but remains wider than that from the CS approach due to the variability introduced in estimating principal stratum probabilities. Using only the first non-fatal event from each subject might seem to diminish the power compared to approaches utilizing all non-fatal events. However, this is not necessarily true, as the PS approach is based on partial likelihood, which yields optimal power. Recurrent event approaches that focus on the marginal feature rely on estimating equations \citep{huang2006analysing}, which exhibit lower efficiency unless the variance structure is correctly specified. \cite{sun2024simple} analyzed COPERNICUS data using all non-fatal events but reported wider CIs. The lack of power may be because risk ratios were derived from estimating equations, which are less efficient than likelihood-based methods. 

The proportionality test is conducted using the method proposed in section B of the Supplementary Materials with $g(t)=t$. The $p$ values are reported in table \ref{copern} and did not reveal violations of the proportionality assumption.

\section{Discussion}\label{c5:sec:discussion}
In clinical trials involving both mortality and morbidity, the cause-specific hazard ratio for the first non-fatal event remains commonly used \citep{buzkova2019assessing} However, it is crucial to acknowledge that this ratio does not reflect the direct effect on the underlying first non-fatal event process, unless the treatment has no differential effects on survival or the population is entirely homogeneous (within each given value of covariates). When treatment has differential effects on survival, the cause-specific hazard ratio should be interpreted with caution. The difference between the cause-specific hazard ratio and the direct effect grows with population heterogeneity, and cardiovascular studies reported significant variability among subjects \citep{cook2016temporal}. \cite{simonetto2022heterogeneity} estimated inter-subject heterogeneity in coronary heart disease risk using mortality data, suggesting that frailty with a variance between 1 and 4 is appropriate to capture total heterogeneity. In our setting, this corresponds to $\Tilde{\gamma}$ between 0.25 and 1. While not directly applicable, their findings offer insights into a plausible range for $\Tilde{\gamma}$ in cardiovascular trials. By specifying a range of $\gamma$, particularly small values, researchers can gauge the potential difference between the cause-specific hazard ratio and the conditional direct effect. In the COPERNICUS trial, the cause-specific hazard ratio is estimated at 0.82, slightly conservative but close to the principal stratum hazard ratio. When utilizing the cause-specific hazard ratio to characterize treatment effects on the first non-fatal event, providing the principal stratum hazard ratio alongside can offer researchers a clearer understanding of the reliability of the cause-specific hazard ratio and aid in understanding the direct effect on the underlying first non-fatal event process.

Over the past decade, it has been argued that for all-cause mortality, the hazard ratio in the Cox model does not provide a causal interpretation when factors influencing the at-risk process are not controlled for. Specifically, the Cox hazard ratio tends to underestimate the causal effect of a beneficial treatment because the active treatment arm often includes relatively more frail subjects compared to the placebo group at each time $t>0$ \citep{hernan2010hazards, aalen2015does, martinussen2020subtleties}. Although the Cox model is never correctly specified in practice (since we cannot control for all factors affecting the at-risk process), it still serves as a useful approximation of the causal effect. Nonetheless, this paper does not focus on providing a single point estimate of the conditional direct effect. Instead, we aim to establish a plausible range for the conditional direct effect, illustrating how the CSHR may differ from the conditional direct effect. This difference arises from the dependence between death and the first non-fatal event, even when conditioned on covariates that might partially explain this dependency. To quantify this difference, we use the frailty approach, tuning the distribution of the frailty according to our beliefs about the extent of this dependency.

The practical application of SACE has faced criticism. For each subject $i$, let $M_i = \min(Y_i^1, Y_i^0)$. For $j = 0, \dots, m$, if $M_i > t_j$, then $i \in \mathcal{A}(t_j)$. Therefore, $\mathcal{A}(0) \supseteq \mathcal{A}(t_1) \supseteq \dots \supseteq \mathcal{A}(t_m)$, where $\mathcal{A}(0)$ represents the entire population, as all subjects can survive at $t=0$. Since $M_i$ is not observable due to its counterfactual nature, it has been argued that the subpopulation $\mathcal{A}(t_j)$, for which the SACE at $t_j$ is relevant, is an unidentifiable subset of the population. This argument holds if we consider the SACE at a specific single $t_j$. However, under the assumption of proportionality, all subjects contribute to the estimation of the principal stratum hazard ratio, unless the subject who experiences a non-fatal event at $t_1$ would not survive to $t_1$ in the counterfactual arm. Therefore, every subject may contribute to the estimation to varying degrees. Additionally, some argue that identifying the SACE relies on untestable assumptions. Despite these criticisms, the value of SACE lies in its role as the only well-defined estimator that measures the direct effect on the underlying first non-fatal event process in the presence of the competing risk of death.

The last known follow-up time for each subject may contribute to estimating the principal stratum probabilities. One approach is to use the frailty $\theta$ to capture the correlation between the first non-fatal event time and potential death times from each treatment arm. This approach implicitly assumes that the correlation between death times is the same as the correlation between death time and the first non-fatal event time. However, the correlation between death times across treatment arms is likely stronger than the correlation between death and the non-fatal event within the same arm. Intuitively, given a death time $Y_i$ from the placebo arm, we expect the potential death time in the active treatment arm to be close to $F_1^{-1}\{ F_0(Y_i) \}$, where $F_1$ and $F_0$ represent the CDF of $Y$. In contrast, uncertainty remains regarding when the first non-fatal event occurs. To accommodate varying degrees of correlations when estimating principal stratum probabilities, one possible approach is to employ a trivariate copula model with two dependence parameters $\varsigma_0$ and $\varsigma_1$, where $\varsigma_1$ measures dependence between $Y^{z}$ and $Y^{1-z}$ and $\varsigma_0$ measures dependence between $T^{z}$ and $Y^{z}$ as well as between $T^{z}$ and $Y^{1-z}$ \citep{zimmer2006using}. Section D of the Supplementary Materials details the estimation of principal stratum probabilities using trivariate copulas and presents the corresponding estimation results for the COPERNICUS trial. Regardless of the method chosen to estimate the principal stratum probabilities, the proposed structure of the PPSH model remains applicable.

In practice, baseline covariates $X$ are typically collected. $X$ can be incorporated into the estimation of principal stratum probabilities. In step 1 outlined in section \ref{membershipest}, baseline covariates can be used to estimate $\hat S_Y(t\mid Z=z,X)$ and $\hat S_T(t\mid Y>t, Z=z, X)$, allowing us to compute $\eta^z_Y(t\mid X)$ and $\eta^z_T(t \mid X)$ in step 2. However, the practical benefit of adjusting for covariates in this context is limited. Including baseline covariates may appear to relax our conditional independence assumptions to $Y^{1-z} \indep T^z \mid X, \theta$ and $Y^{z} \indep T^z \mid X, \theta$, implying that $X$ accounts for some of the correlation between potential outcomes. Nevertheless, we pre-specify a range of $\gamma$ values, and the correlation strength implied by this range generally covers what can be explained by $X$. Thus, covariate adjustment in the estimation of principal stratum probabilities may be redundant. When fitting the PPSH model, covariates $X$ can be included in equation (\ref{psphform}) under a multiplicative effect assumption, similar to the Cox model. In this case, the interpretation of the principal stratum hazard ratio shifts from marginal to conditional due to non-collapsibility. Section D of the Supplementary Materials presents the estimated principal stratum hazard ratios, conditional on baseline covariates, for the COPERNICUS trial. 

The PPSH model is designed to address the time to the first non-fatal event, wherein subjects who experience the non-fatal event are no longer at risk for that event. However, there is a possibility of expanding our framework to recurrent event scenarios, similar to the Anderson-Gill model, which serves as an extension of the Cox model.

\bibliography{maintext}  

\clearpage

\begin{table}[htbp]
\centering
\scriptsize
\caption{Comparison of estimation results between the principal stratum and the cause-specific approaches. \label{results}}
{\tabcolsep=3.5pt
\begin{tabular}{cccccccccccccccccc}
\toprule
& \multicolumn{10}{c}{Summary Statistics} & \multicolumn{7}{c}{Estimation of $\beta$} \\
\cmidrule(lr){2-11} \cmidrule(lr){12-18}
& \multicolumn{5}{c}{Placebo} & \multicolumn{5}{c}{Active Treatment} & \multicolumn{2}{c}{Hypothetical} & \multicolumn{2}{c}{Cause-Specific} & \multicolumn{3}{c}{Principal Stratum} \\ 
\cmidrule(lr){2-6} \cmidrule(lr){7-11} \cmidrule(lr){12-13} \cmidrule(lr){14-15} \cmidrule(lr){16-18}
$\gamma$ & Dead & Censored & LOF & ``D'' & Event & Dead & Censored & LOF & ``D'' & Event & Bias (SE) & HR & Bias (SE) & HR & $\Tilde{\gamma}$ & Bias (SE) & HR \\ \midrule
\multicolumn{18}{c}{$\mathbf{\lambda_0 = 0.25}$} \\
\multirow{3}{*}{0.5} 
& \multirow{3}{*}{29\%} & \multirow{3}{*}{67\%} & \multirow{3}{*}{5\%} & \multirow{3}{*}{1.6} & \multirow{3}{*}{60\%} 
& \multirow{3}{*}{25\%} & \multirow{3}{*}{70\%} & \multirow{3}{*}{5\%} & \multirow{3}{*}{1.7} & \multirow{3}{*}{39\%} 
& \multirow{3}{*}{0.002 (0.005)} & \multirow{3}{*}{0.5} & \multirow{3}{*}{0.054 (0.005)} & \multirow{3}{*}{0.53} & 0.5 & -0.003 (0.006) & 0.50 \\
& & & & & & & & & & & & & & & 2.0 & 0.039 (0.006) & 0.52 \\
& & & & & & & & & & & & & & & 5.0 & 0.046 (0.005) & 0.52 \\[1ex]

\multirow{3}{*}{2.0} 
& \multirow{3}{*}{35\%} & \multirow{3}{*}{60\%} & \multirow{3}{*}{5\%} & \multirow{3}{*}{1.6} & \multirow{3}{*}{80\%} 
& \multirow{3}{*}{30\%} & \multirow{3}{*}{65\%} & \multirow{3}{*}{5\%} & \multirow{3}{*}{1.6} & \multirow{3}{*}{58\%} 
& \multirow{3}{*}{-0.003 (0.004)} & \multirow{3}{*}{0.5} & \multirow{3}{*}{0.008 (0.005)} & \multirow{3}{*}{0.50} & 0.5 & -0.044 (0.005) & 0.48 \\
& & & & & & & & & & & & & & & 2.0 & -0.010 (0.005) & 0.50 \\
& & & & & & & & & & & & & & & 5.0 & 0.001 (0.005) & 0.50 \\ [1ex]

\multirow{3}{*}{5.0} 
& \multirow{3}{*}{37\%} & \multirow{3}{*}{58\%} & \multirow{3}{*}{5\%} & \multirow{3}{*}{1.5} & \multirow{3}{*}{84\%} 
& \multirow{3}{*}{31\%} & \multirow{3}{*}{64\%} & \multirow{3}{*}{5\%} & \multirow{3}{*}{1.6} & \multirow{3}{*}{67\%} 
& \multirow{3}{*}{-0.001 (0.004)} & \multirow{3}{*}{0.5} & \multirow{3}{*}{0 (0.004)} & \multirow{3}{*}{0.50} & 0.5 & -0.045 (0.004) & 0.48 \\
& & & & & & & & & & & & & & & 2.0 & -0.018 (0.004) & 0.49 \\
& & & & & & & & & & & & & & & 5.0 & -0.007 (0.004) & 0.50 \\ [1ex]

\multicolumn{18}{c}{$\mathbf{\lambda_0 = 0.4}$} \\
\multirow{3}{*}{0.5} 
& \multirow{3}{*}{37\%} & \multirow{3}{*}{58\%} & \multirow{3}{*}{4\%} & \multirow{3}{*}{1.5} & \multirow{3}{*}{57\%} 
& \multirow{3}{*}{25\%} & \multirow{3}{*}{70\%} & \multirow{3}{*}{5\%} & \multirow{3}{*}{1.7} & \multirow{3}{*}{40\%} 
& \multirow{3}{*}{0.002 (0.005)} & \multirow{3}{*}{0.5} & \multirow{3}{*}{0.107 (0.005)} & \multirow{3}{*}{0.56} & 0.5 & 0.002 (0.006) & 0.50 \\
& & & & & & & & & & & & & & & 2.0 & 0.071 (0.006) & 0.54 \\
& & & & & & & & & & & & & & & 5.0 & 0.088 (0.005) & 0.55 \\ [1ex]

\multirow{3}{*}{2.0} 
& \multirow{3}{*}{48\%} & \multirow{3}{*}{48\%} & \multirow{3}{*}{4\%} & \multirow{3}{*}{1.4} & \multirow{3}{*}{75\%} 
& \multirow{3}{*}{30\%} & \multirow{3}{*}{65\%} & \multirow{3}{*}{5\%} & \multirow{3}{*}{1.6} & \multirow{3}{*}{59\%} 
& \multirow{3}{*}{-0.003 (0.004)} & \multirow{3}{*}{0.5} & \multirow{3}{*}{0.034 (0.005)} & \multirow{3}{*}{0.52} & 0.5 & -0.067 (0.005) & 0.47 \\
& & & & & & & & & & & & & & & 2.0 & -0.008 (0.005) & 0.50 \\
& & & & & & & & & & & & & & & 5.0 & 0.015 (0.005) & 0.51 \\ [1ex]

\multirow{3}{*}{5.0} 
& \multirow{3}{*}{51\%} & \multirow{3}{*}{45\%} & \multirow{3}{*}{4\%} & \multirow{3}{*}{1.4} & \multirow{3}{*}{80\%} 
& \multirow{3}{*}{31\%} & \multirow{3}{*}{64\%} & \multirow{3}{*}{5\%} & \multirow{3}{*}{1.6} & \multirow{3}{*}{67\%} 
& \multirow{3}{*}{-0.001 (0.004)} & \multirow{3}{*}{0.5} & \multirow{3}{*}{0.012 (0.004)} & \multirow{3}{*}{0.51} & 0.5 & -0.078 (0.005) & 0.46 \\
& & & & & & & & & & & & & & & 2.0 & -0.030 (0.004) & 0.49 \\
& & & & & & & & & & & & & & & 5.0 & -0.006 (0.004) & 0.50 \\
\bottomrule
\end{tabular}}
\begin{tablenotes}
\item[] \textbf{Notes:}
\begin{itemize}
\item $\gamma$ is the inverse variance of gamma-distributed frailty. 
\item ``Dead'' refers to the average proportions of death.
\item ``Censored'' refers to the average proportions of administrative censoring.
\item ``LOF'' refers to the average proportion of loss of follow-up.
\item ``D'' refers to the mean duration of the follow-up times.
\item ``Event'' refers to the average proportion of patients with a non-fatal event.
\item $\Tilde{\gamma}$ is the pre-specified value used in the principal stratum estimation.
\end{itemize}
\end{tablenotes}
\end{table}

\begin{table}[htbp]
\centering
\scriptsize
\caption{Sensitivity analysis comparing the principal stratum and cause-specific approaches.\label{sensitivityresults}}
{\tabcolsep=3.5pt
\begin{tabular}{cccccccccccccccccc}
\toprule
& \multicolumn{10}{c}{Summary Statistics} & \multicolumn{7}{c}{Estimation of $\beta$} \\
\cmidrule(lr){2-11} \cmidrule(lr){12-18}
& \multicolumn{5}{c}{Placebo} & \multicolumn{5}{c}{Active Treatment} & \multicolumn{2}{c}{Hypothetical} & \multicolumn{2}{c}{Cause-Specific} & \multicolumn{3}{c}{Principal Stratum} \\ 
\cmidrule(lr){2-6} \cmidrule(lr){7-11} \cmidrule(lr){12-13} \cmidrule(lr){14-15} \cmidrule(lr){16-18}
$\gamma$ & Dead & Censored & LOF & ``D'' & Event & Dead & Censored & LOF & ``D'' & Event & Est (SE) & HR & Bias (SE) & HR & $\Tilde{\gamma}$ & Bias (SE) & HR \\ \midrule
\multicolumn{18}{c}{$\mathbf{\lambda_0 = 0.25}$} \\
\multirow{3}{*}{0.5} & \multirow{3}{*}{30\%} & \multirow{3}{*}{65\%} & \multirow{3}{*}{5\%} & \multirow{3}{*}{1.6} & \multirow{3}{*}{71\%} & \multirow{3}{*}{26\%} & \multirow{3}{*}{69\%} & \multirow{3}{*}{5\%} & \multirow{3}{*}{1.7} & \multirow{3}{*}{43\%} & \multirow{3}{*}{-0.887 (0.005)} & \multirow{3}{*}{0.41} & \multirow{3}{*}{0.078 (0.005)} & \multirow{3}{*}{0.45} & 0.5 & 0.016 (0.006) & 0.42 \\
&&&&&&&&&&&&&&& 2.0 & 0.070 (0.005) & 0.44 \\
&&&&&&&&&&&&&&& 5.0 & 0.081 (0.005) & 0.45 \\[1ex]
\multirow{3}{*}{2.0} & \multirow{3}{*}{35\%} & \multirow{3}{*}{60\%} & \multirow{3}{*}{5\%} & \multirow{3}{*}{1.6} & \multirow{3}{*}{82\%} & \multirow{3}{*}{30\%} & \multirow{3}{*}{65\%} & \multirow{3}{*}{5\%} & \multirow{3}{*}{1.6} & \multirow{3}{*}{60\%} & \multirow{3}{*}{-0.748 (0.004)} & \multirow{3}{*}{0.47} & \multirow{3}{*}{0.020 (0.005)} & \multirow{3}{*}{0.48} & 0.5 & -0.032 (0.005) & 0.46 \\
&&&&&&&&&&&&&&& 2.0 & 0.004 (0.005) & 0.48 \\
&&&&&&&&&&&&&&& 5.0 & 0.017 (0.005) & 0.48 \\[1ex]
\multirow{3}{*}{5.0} & \multirow{3}{*}{37\%} & \multirow{3}{*}{58\%} & \multirow{3}{*}{5\%} & \multirow{3}{*}{1.5} & \multirow{3}{*}{85\%} & \multirow{3}{*}{31\%} & \multirow{3}{*}{64\%} & \multirow{3}{*}{5\%} & \multirow{3}{*}{1.6} & \multirow{3}{*}{67\%} & \multirow{3}{*}{-0.713 (0.004)} & \multirow{3}{*}{0.49} & \multirow{3}{*}{0.009 (0.004)} & \multirow{3}{*}{0.49} & 0.5 & -0.036 (0.005) & 0.47 \\
&&&&&&&&&&&&&&& 2.0 & -0.008 (0.004) & 0.49 \\
&&&&&&&&&&&&&&& 5.0 & 0.004 (0.004) & 0.49 \\[1ex]
\multicolumn{18}{c}{$\mathbf{\lambda_0 = 0.4}$} \\
\multirow{3}{*}{0.5} & \multirow{3}{*}{40\%} & \multirow{3}{*}{56\%} & \multirow{3}{*}{4\%} & \multirow{3}{*}{1.5} & \multirow{3}{*}{68\%} & \multirow{3}{*}{26\%} & \multirow{3}{*}{69\%} & \multirow{3}{*}{5\%} & \multirow{3}{*}{1.7} & \multirow{3}{*}{44\%} & \multirow{3}{*}{-0.887 (0.005)} & \multirow{3}{*}{0.41} & \multirow{3}{*}{0.124 (0.005)} & \multirow{3}{*}{0.47} & 0.5 & 0.009 (0.006) & 0.42 \\
&&&&&&&&&&&&&&& 2.0 & 0.095 (0.005) & 0.45 \\
&&&&&&&&&&&&&&& 5.0 & 0.116 (0.005) & 0.46 \\[1ex]
\multirow{3}{*}{2.0} & \multirow{3}{*}{48\%} & \multirow{3}{*}{48\%} & \multirow{3}{*}{4\%} & \multirow{3}{*}{1.4} & \multirow{3}{*}{77\%} & \multirow{3}{*}{30\%} & \multirow{3}{*}{65\%} & \multirow{3}{*}{5\%} & \multirow{3}{*}{1.6} & \multirow{3}{*}{60\%} & \multirow{3}{*}{-0.748 (0.004)} & \multirow{3}{*}{0.47} & \multirow{3}{*}{0.040 (0.005)} & \multirow{3}{*}{0.49} & 0.5 & -0.061 (0.005) & 0.45 \\
&&&&&&&&&&&&&&& 2.0 & 0.002 (0.005) & 0.47 \\
&&&&&&&&&&&&&&& 5.0 & 0.026 (0.005) & 0.49 \\[1ex]
\multirow{3}{*}{5.0} & \multirow{3}{*}{51\%} & \multirow{3}{*}{45\%} & \multirow{3}{*}{4\%} & \multirow{3}{*}{1.4} & \multirow{3}{*}{80\%} & \multirow{3}{*}{31\%} & \multirow{3}{*}{64\%} & \multirow{3}{*}{5\%} & \multirow{3}{*}{1.6} & \multirow{3}{*}{68\%} & \multirow{3}{*}{-0.713 (0.004)} & \multirow{3}{*}{0.49} & \multirow{3}{*}{0.019 (0.004)} & \multirow{3}{*}{0.50} & 0.5 & -0.071 (0.005) & 0.46 \\
&&&&&&&&&&&&&&& 2.0 & -0.021 (0.005) & 0.48 \\
&&&&&&&&&&&&&&& 5.0 & 0.003 (0.004) & 0.49 \\
\bottomrule
\end{tabular}}
\begin{tablenotes}
\item[] \textbf{Notes:}
\begin{itemize}
\item $\gamma$ is the inverse variance of inverse Gaussian distributed frailty. 
\item ``Dead'' refers to the average proportions of death. 
\item ``Censored'' refers to the average proportions of administrative censoring. 
\item ``LOF'' refers to the average proportion of loss of follow-up. 
\item ``D'' refers to the mean duration of the follow-up times. 
\item ``Event'' refers to the average proportion of patients with a non-fatal event. 
\item $\Tilde{\gamma}$ is the pre-specified value used in the principal stratum estimation.
\end{itemize}
\end{tablenotes}
\end{table}

\begin{table}[htbp]
\centering
\caption{Estimation results for the COPERNICUS trial data. \label{copern}}
{\tabcolsep=25pt
\begin{tabular}{ccccc}
\toprule
Approach              & $\Tilde{\gamma}$ & HR & 95\% CI & Proportionality \\ \midrule

\multirow{6}{*}{PS} & 0.25   & 0.781   & (0.665, 0.919) & $p=0.64$  \\ 
                    & 0.5    & 0.791   & (0.676, 0.927) & $p=0.61$ \\
                    & 1      & 0.804   & (0.692, 0.934) & $p=0.68$ \\
                    & 2      & 0.813   & (0.702, 0.941) & $p=0.76$ \\
                    & 5      & 0.818   & (0.708, 0.948) & $p=0.81$ \\
                    & 10     & 0.820  &  (0.711, 0.949) & $p=0.84$ \\ \midrule
CS                  & $\infty$        & 0.821   & (0.716, 0.943) & -  \\           
\bottomrule
\end{tabular}}
\end{table}

\end{document}


\maketitle

\section{Technical Issues in Estimating \texorpdfstring{$\gamma$}{gamma}}
If we use $\theta \sim \Gamma(\gamma,\gamma)$ to induce correlation between the observed first non-fatal event and the observed death time, we could formulate a likelihood function to estimate $\gamma$. The likelihood function involves $\gamma$ and conditional cumulative hazard functions. If we empirically estimate the conditional cumulative hazard from the marginal distribution, we obtain the profile likelihood, which solely involves $\gamma$. 

However, the profile log-likelihood does not adhere to the conventional sense of a log-likelihood function, and its derivative does not exhibit a zero mean, a crucial property for estimating equations. The profile likelihood is satisfactory when the dimension of $\eta^{z}_{Y}(t)$, $\eta^{z}_{T}(t)$, and $\eta^{1-z}_{Y}(t)$ is small relative to the total Fisher information \citep{mccullagh2019generalized} but unfortunately, this is not the case for the derived profile likelihood on $\gamma$, which yields unsatisfactory results. 

Given the limitations of the profile likelihood, one solution is to assume parametric distributions for the first non-fatal event time and death time. Subsequently, the likelihood function incorporates $\gamma$ and parameters governing the non-fatal event and death processes. However, the likelihood function on $\gamma$ is observed to be bimodal: the first mode centers around the true value of $\gamma$ while the second mode appears at a very large value. This bimodality leads to unstable estimation and, in some cases, unrealistically large estimates of $\gamma$.

\section{Testing Proportionality}\label{app:prop}
A key assumption of the PPSH model is proportionality. This assumption states that, with time-fixed covariates, the relative principal stratum hazard for any two subjects $i$ and $j$ follows a constant ratio:
$$ \frac{\hbox{exp}(\beta Z_i) \lambda^{P}_{0}(t)}{\hbox{exp}(\beta Z_j) \lambda^{P}_{0}(t)}=\frac{\hbox{exp}(\beta Z_i)}{\hbox{exp}(\beta Z_j)}$$ which is independent of time. 

The Schoenfeld residuals \citep{schoenfeld1982partial} are particularly useful for detecting violations of the proportionality assumption. These residuals are defined only for those subjects with a non-fatal event and are essentially the subject-level contributions to the modified partial likelihood score function. At each $t_j \in \mathcal{S}$, let 
$$\bar Z(t_j; \beta) =  \frac{\sum_{i \in \mathcal{R}_j} p_{ij}  Z_{i}  \hbox{exp}(\beta Z_{i})}{\sum_{i \in \mathcal{R}_j} p_{ij}  \hbox{exp}(\beta Z_{i})}$$
which is the weighted covariate mean. The Schoenfeld residual at $t_j \in \mathcal{S}$ is defined as:
\begin{eqnarray}
    s_{j}= p_{(j)j}  \{ Z_{(j)}-\bar Z(t_j; \beta) \} .\label{scho}
\end{eqnarray}

When there are $k>1$ tied events at a given time $t_j$, we compute $k$ residuals, each one based on equation (\ref{scho}) applied to that event rather than an overall residual for that time point. 

The proportionality assumption can be relaxed by allowing time-varying $\beta(t)$. Suppose 
\begin{eqnarray}
    \beta(t)=\beta + \xi  g(t) \label{tvarybeta}
\end{eqnarray}
where $g(t)$ is a transformation of the time scale. Under this formulation, testing for proportionality is equivalent to testing whether $\xi$ equals zero.

Under model (\ref{tvarybeta}), equation (\ref{scho}) can be written as the sum of the Schoenfeld residuals from the model (\ref{tvarybeta}) with a mean 0, and the difference between the weighted covariate means. That is,
\begin{eqnarray*}
    (\ref{scho}) = p_{(j)j}  \left( [ Z_{(j)} -\bar Z\{t_j; \beta(t_j)\} ] + [\bar Z\{t_j; \beta(t_j)\} -\bar Z(t_j; \beta)] \right).
\end{eqnarray*}

Expanding $\bar Z\{t_j; \beta(t_j)\}$ in a one-term Taylor expansion about $\beta(t_j)=\beta$ under model (\ref{tvarybeta}), we have $\bar Z\{t_j; \beta(t_j)\} \approx \{p_{(j)j}\}^{-1}  V(t_j;\beta)  \{\beta(t)-\beta\}=\{p_{(j)j}\}^{-1}  V(t_j;\beta)  \xi  g(t_j)$. Let $s^{*}_j=V(t_j;\beta)^{-1}  s_j$ be the scaled Schoenfeld residual. We have
\begin{eqnarray}
\hbox{E}(s^{*}_j) = V(t_j;\beta)^{-1}  \hbox{E}(s_j) \approx V(t_j;\beta)^{-1}  V(t_j;\beta)  \xi g(t_j) = \xi g(t_j) \label{scaledscho_mean}.
\end{eqnarray}

Model (\ref{tvarybeta}) and equation (\ref{scaledscho_mean}) suggest plotting $s^{*}_j$ versus $g(t)$ as a visual assessment for proportionality. A regression line can be fitted to the plot, followed by a test for $\xi=0$; a non-zero slope suggests evidence against proportionality. Additionally, $g(t)$ can be estimated by creating a smoothed plot of $s^{*}_j$ against $t_j$.

To formally derive the proportionality test, we first need to obtain the variance of $\hat s^{*}_j \coloneqq s^{*}_{j}(\hat \beta)$, which can be calculated from the variance of $\hat s_j \coloneqq s_{j}(\hat \beta)$. While $s_j$'s are uncorrelated, their estimates $\hat s_j$'s are correlated because $\sum_{j:t_j \in \mathcal{S}} \hat s_j =0$.

We can expand $\hat s_{j}$ in a Taylor series:
\begin{eqnarray*}
    \hat s_{j} \approx s_{j} - \widehat V_j (\hat \beta -\beta) \approx s_{j} - \widehat V_j  \mathcal{L}(\hat \beta)^{-1} \sum_{j: t_j \in \mathcal{S}}s_j
\end{eqnarray*}
where $\hat \beta -\beta$ is substituted by the score statistic, and $\widehat V_j=V(t_j; \hat \beta)$. Then 
\begin{eqnarray*}
    \hbox{Var}(\hat s_{j}) \approx \widehat V_j + \widehat V_j  \mathcal{L}(\hat \beta)^{-1} \sum_{j: t_j \in \mathcal{S}} V_j  \mathcal{L}(\hat \beta)^{-1}  \widehat V_j - 2 \widehat V_j  \mathcal{L}(\hat \beta)^{-1}  \widehat V_j = \widehat V_j - \widehat V_j \mathcal{L}(\hat \beta)^{-1} \widehat V_j.
\end{eqnarray*}
Then
\begin{eqnarray}
    \hbox{Var}(\hat s^{*}_{j}) \approx \widehat V_j^{-1} - \mathcal{L}(\hat \beta)^{-1} \to \widehat V_j^{-1}  \label{scaledscho_var}
\end{eqnarray}
where the approximation is satisfactory if $\widehat V_j$ is a small proportion of $\mathcal{L}(\hat \beta)=\sum_{j: t_j \in \mathcal{S}}\widehat V_j$, which is almost always the case unless there are only a few event times.

Equations (\ref{scaledscho_mean}) and (\ref{scaledscho_var}) suggest a standard linear model for $\hat s^{*}_{j}$: $\hat s^{*}_{j}= g(t_j)  \xi  + \epsilon$. Generalized least squares gives 
\begin{eqnarray*}
    \hat \xi = \left\{ \sum_{j: t_j \in \mathcal{S}} g(t_j)  \widehat V_j  g(t_j) \right\}^{-1} \sum_{j: t_j \in \mathcal{S}} g(t_j)  \hat s_{j} .
\end{eqnarray*}

The variance of $\hat \xi$ can be consistently estimated by $\left\{ \sum_{j: t_j \in \mathcal{S}} g(t_j)  \widehat V_j  g(t_j) \right\}^{-1}$ under the null hypothesis that $\xi=0$, leading to an asymptotic $\chi^{2}$ test statistic of the null hypothesis on 1 degree of freedom:
\begin{eqnarray*}
    \left\{ \sum_{j: t_j \in \mathcal{S}} g(t_j)  \hat s_{j}  \right\} \left\{ \sum_{j: t_j \in \mathcal{S}} g(t_j)  \widehat V_j  g(t_j) \right\}^{-1} \left\{ \sum_{j: t_j \in \mathcal{S}} g(t_j)  \hat s_{j}  \right\} \sim \chi^{2}_{1} .
\end{eqnarray*}

The idea behind the proportionality test described in this section is not new; it originates from \cite{grambsch1994proportional}. They also demonstrated that if $g(t)$ is a specified function of time, then the proposed test is analogous to a score test for incorporating the time-dependent variable $g(t)Z$ into the model.

\section{\texorpdfstring{Derivation of $\eta^{1}_{T}(t)$ in the Simulation}{Derivation of eta	extunderscore T	extasciicircum 1(t) in the Simulation}}\label{app:cumuhazard}
In our simulation setting, we make the following assumptions: (1) independent loss to follow-up, denoted by $C$, (2) the frailty $\theta$ follows a gamma distribution with a mean of one and a variance of $1/\gamma$, and (3) mutual independence between $(Y^z, Y^{1-z}, T^{z})$ conditional on $\theta$. Then, 
\begin{eqnarray*}
    \lambda^{P}(t \mid Z=z) \hbox{d}t &=& P(t \leq T^{z} < t+dt \mid T^{z} \geq t, Y^{z} > t, Y^{1-z} > t, C > t) \\
    &=& P(t \leq T^{z} < t+dt \mid T^{z} \geq t, Y^{z} > t, Y^{1-z} > t) \\
    &=& \frac{\int P(t \leq T^{z} < t+dt \mid \theta)  P(Y^{z}> t \mid \theta)  P(Y^{1-z}> t \mid \theta)  p(\theta)d\theta}{\int P(T^{z} \geq t \mid \theta)  P(Y^{z} > t \mid \theta)  P(Y^{1-z} > t \mid \theta)  p(\theta)d\theta} \\
    &=& \frac{\int \theta  d\eta^{z}_{T}(t)  \hbox{exp}\{-\theta \eta^{z}_{T}(t)\}  \hbox{exp}(-\theta \lambda_0 t)  \hbox{exp}(-\theta \lambda_1 t)  \frac{\gamma^{\gamma}}{\Gamma(\gamma)} \theta^{\gamma -1} \hbox{exp}(-\gamma \theta) d\theta}{\int \hbox{exp}\{-\theta \eta^{z}_{T}(t) \} \hbox{exp}(-\theta \lambda_0 t)  \hbox{exp}(-\theta \lambda_1 t)  \frac{\gamma^{\gamma}}{\Gamma(\gamma)} \theta^{\gamma -1} \hbox{exp}(-\gamma \theta) d\theta}  \\
    &=& d\eta^{z}_{T}(t)  \frac{ \left( \frac{\gamma}{\gamma + \eta^{z}_{T}(t) + \lambda_0 t + \lambda_1 t} \right)^{\gamma +1}}{\left( \frac{\gamma}{\gamma + \eta^{z}_{T}(t) + \lambda_0 t + \lambda_1 t} \right)^{\gamma}} = d\eta^{z}_{T}(t)  \frac{\gamma}{\gamma + \eta^{z}_{T}(t) + \lambda_0 t + \lambda_1 t} 
\end{eqnarray*}
where $dt$ and $d\eta^{z}_{T}(t)$ represents an infinitesimally small change in $t$ and $\eta^{z}_{T}(t)$, respectively.

The marginal proportional PSHR of $r^P$ is then given by:
$$r^P=\frac{\lambda^{P}(t \mid Z=1)}{\lambda^{P}(t \mid Z=0)}= \frac{\gamma + \phi t + \lambda_0 t + \lambda_1 t}{\gamma + \eta^{1}_{T}(t) + \lambda_0 t + \lambda_1 t}  \frac{1}{\phi}  \frac{d\eta^{1}_{T}(t)}{dt} .$$

Let $y=\eta^{1}_{T}(t)$, and rearranging the above equation, we obtain:
$$\frac{dy}{dt}+ P(t)  y = Q(t)$$
where 
$$P(t)=-\frac{r^P\phi}{\gamma+\phi t + \lambda_0 t + \lambda_1 t},$$ $$Q(t)=r^P\phi  \frac{\gamma + \lambda_0 t + \lambda_1 t}{\gamma+\phi t + \lambda_0 t + \lambda_1 t}.$$

Now, the problem is reduced to solving a non-homogeneous linear first-order differential equation.

\textbf{Step 1:} Let $Q(t)=0$ and solve the complementary equation $dy/dt+P(t)y=0$. Then
$$y=U  \hbox{exp}\left\{-\int P(t)dt\right\}=U  \left\{ \gamma + (\phi + \lambda_0 + \lambda_1)t  \right\}^{\frac{r^P\phi}{\phi + \lambda_0 + \lambda_1}}$$ where $U$ is a constant.

\textbf{Step 2:} Now, we replace $U$ with a certain function $u(t)$ and will find a solution to the original non-homogeneous equation. Let $y=u(t)  \hbox{exp}\left\{-\int P(t)dt\right\}$, then 
$$Q(t)=\frac{dy}{dt}+P(t)y=\frac{du(t)}{dt}  \hbox{exp}\left\{-\int P(t)dt\right\}.$$

Then
\begin{eqnarray*}
    u(t) &=& \int \frac{Q(t)}{\hbox{exp}\left\{-\int P(t)dt\right\}}dt + U^{*} \\
    &=& \phi  \frac{\gamma - \gamma r^P - (\lambda_0 +\lambda_1)r^Pt}{\{ \phi(r^P-1)-\lambda_0 -\lambda_1 \}  \{ \gamma + (\phi + \lambda_0 + \lambda_1)t \}^{\frac{r^P\phi}{\phi + \lambda_0 + \lambda_1}}} + U^{*}
\end{eqnarray*}
where $U^{*}$ is a constant.

Then the general solution is written in the form:
\begin{eqnarray*}
   y&=&u(t)  \hbox{exp}\left\{-\int P(t)dt\right\}= u(t)  \left\{ \gamma + (\phi + \lambda_0 + \lambda_1)t  \right\}^{\frac{r^P\phi}{\phi + \lambda_0 + \lambda_1}} \\
   &=& U^{*}  \left\{ \gamma + (\phi + \lambda_0 + \lambda_1)t  \right\}^{\frac{r^P\phi}{\phi + \lambda_0 + \lambda_1}} + \phi  \frac{\gamma - \gamma r^P - (\lambda_0 +\lambda_1)r^Pt}{ \phi(r^P-1)-\lambda_0 -\lambda_1  }.
\end{eqnarray*}

\textbf{Step 3:} 
To determine $U^{*}$, we observe that at $t=0$, $y(0)=0$, then 
$$0=U^{*}  \gamma^{\frac{r^P\phi}{\phi + \lambda_0 + \lambda_1}} + \phi  \frac{\gamma - \gamma r^P}{ \phi(r^P-1)-\lambda_0 -\lambda_1  }.$$

We obtain
$$U^{*}=\frac{\phi (1-r^P)}{\lambda_0 + \lambda_1 - \phi(r^P-1)}  \gamma^{1-\frac{r^P\phi}{\phi + \lambda_0 + \lambda_1 }}.$$

Then
\begin{eqnarray*}
    \eta^{1}_{T}(t)&=&y \\ 
    &=& \frac{\phi}{\lambda_{Y}-\phi(r^P-1)} \left[ (1-r^P)  \gamma^{1-\frac{r^P\phi}{\phi + \lambda_{Y}}}  \{ \gamma + (\phi + \lambda_{Y})t \}^{\frac{r^P\phi}{\phi + \lambda_{Y}}} - \gamma +\gamma r^P + \lambda_{Y}r^Pt \right]
\end{eqnarray*}
where $\lambda_{Y}=\lambda_{0}+\lambda_{1}$.

\section{Accommodating Varying Degrees of Correlations Using Trivariate Copulas}\label{app:copula}

\subsection{Extension of Bivariate Copulas to Trivariate}
\subsubsection{Bivariate Copulas}
Let $E^z=\hbox{min}(Y^z, T^z)$ denote the potential outcome for the first event time $E$ under treatment arm $z$. The joint survival function of $E^z$ and $Y^z$ can be expressed using their univariate marginal survival functions and a copula that captures the correlation structure between the two variables \citep{sklar1959fonctions}. For $n$ pairs of realization of $E^z$ and $Y^z$, denoted as $(e^z_i, y^z_i)$ for $1 \leq i \leq n$, let $\delta_{1i}$ and $\delta_{2i}$ represent the corresponding status indicators. The indicator $\delta_{1i}$ equals 1 if $e^z_i$ is observed and 0 if censored, while $\delta_{2i}$ is defined analogously for $y^z_i$. It is worth noting that, due to the definition $E^z = \min(Y^z, T^z)$, it is impossible for the death $Y^z$ to occur while the first event $E^z$ does not (e.g., $\delta_{1i} = 0$ and $\delta_{2i} = 1$). Let $u_{1i}=S_{E}^{z}(e^z_i)$ and $u_{2i}=S_{Y}^{z}(y^z_i)$ represent the IID realizations of $U_1$ and $U_2$, where $S(\cdot)$ denotes the survival function. The joint survival function of $E^z$ and $Y^z$ can then be expressed as $\mathbb{C}_{\varsigma}(U_1, U_2)$, where $\mathbb{C}$, the copula, is a multivariate survival function for which the marginal survival functions of $U_1$ and $U_2$ are both uniform over the interval $[0,1]$. There are four commonly used types of copulas:
\begin{itemize}
    \item Clayton: $\mathbb{C}_{\varsigma}(U_1, U_2)=(U_1^{-\varsigma} + U_2^{-\varsigma} -1)^{-1/\varsigma}$
    \item Gumbel: $\mathbb{C}_{\varsigma}(U_1, U_2)=\hbox{exp}\left[ - \left\{(-\hbox{log}U_1)^{\varsigma} +(-\hbox{log}U_2)^{\varsigma}   \right\}^{1/\varsigma} \right] $
    \item Frank: $\mathbb{C}_{\varsigma}(U_1, U_2)=-\frac{1}{\varsigma}\hbox{log} \left[ 1+ \frac{ \left\{\hbox{exp}(-\varsigma U_1)-1\right\}\left\{\hbox{exp}(-\varsigma U_2)-1\right\} }{\hbox{exp}(-\varsigma)-1} \right] $
    \item Normal: $\mathbb{C}_{\varsigma}(U_1, U_2)=\int_{-\infty}^{\Phi^{-1}(U_1)} \int_{-\infty}^{\Phi^{-1}(U_2)} \frac{1}{2\pi \sqrt{1-\varsigma^2}} \hbox{exp}\left\{ -\frac{s_1^2 - 2\varsigma s_1 s_2 + s_2^2}{2(1-\varsigma^2)} \right\} ds_1 ds_2  $, where $\Phi^{-1}(\cdot)$ is the inverse CDF of a standard normal.
\end{itemize} 

The parameter $\varsigma$ in each copula reflects the degree of rank correlation. An important property of ranks is their invariance under strictly monotonic transformations. Specifically, transforming each variable by its CDF does not alter ranks, implying that the distribution of any rank-based statistic depends solely on the copula, not on the univariate marginals \citep{joe2014dependence}. In biomedical applications, Kendall's tau, $\tau$, is a commonly used rank statistic. For uncensored data pairs $(e^z_i, y^z_i)$ and $(e^z_j, y^z_j)$, where $\delta_{1i}=1, \delta_{2i}=1, \delta_{1j}=1, \delta_{2j}=1$, Kendall's tau is defined as: $\tau=P\{ (e^z_i- e^z_j)(y^z_i - y^z_j)>0 \} - P\{ (e^z_i- e^z_j)(y^z_i - y^z_j)<0 \}$, which represents the probability of a concordant pair minus the probability of a discordant pair. When censoring is present, several methods have been proposed for nonparametric estimation of Kendall's tau. These include replacing censored event times with appropriate imputations \citep{hsieh2010estimation}, using modified estimators with Inverse Probability Censoring Weighting \citep{lakhal2009ipcw}, and potentially other approaches. For Clayton, Gumbel, and Frank copulas, the relationship between $\varsigma$ and Kendall's tau is given by $\tau=\varsigma/(2+\varsigma)$, $\tau=1-1/\varsigma$, and $\tau=1+4\{ \frac{1}{\varsigma} \int_{0}^{\varsigma} \frac{t}{\hbox{exp}(t)-1} dt -1 \}/\varsigma$, respectively. 

\subsubsection{Trivariate Extensions}
The Clayton, Gumbel, and Frank copulas belong to the broader family of Archimedean copulas. The bivariate Archimedean copulas take the general form $\mathbb{C}_{\varsigma}(U_1, U_2)=\Psi\{\Psi^{-1}(U_1)+\Psi^{-1}(U_2)\}$, where $\Psi(U)$ is a generating function that satisfies certain conditions. The joint survival function of $E^z$, $Y^z$, and $Y^{1-z}$ can also be expressed in terms of univariate marginal survival functions and a trivariate copula. The trivariate copulas can be constructed from some bivariate Archimedean copulas using the mixtures-of-power approach \citep{zimmer2006using}, which employs the Laplace transformation as a special generator and produces two dependence parameters. The Clayton, Gumbel, and Frank copulas can be extended to trivariate cases by expressing them in terms of mixtures-of-powers. Specifically, the trivariate extension using the mixtures-of-power approach is:
\begin{eqnarray*}
    \mathbb{C}_{\varsigma_0, \varsigma_1}(U_1, U_2, U_3)&=&\Psi_0[\Psi_0^{-1}(U_1) + \Psi_0^{-1} \circ \Psi_1 \{ \Psi_1^{-1}(U_2) + \Psi_1^{-1}(U_3) \} ] \\
    &=& \mathbb{C}\{U_1, \mathbb{C}(U_2, U_3; \Psi_1); \Psi_0\}
\end{eqnarray*}
where $f \circ g$ denotes the functional operation $f\{g(x) \}$, and $u_{3i}=S_{Y}^{1-z}(y^{1-z}_{i})$ are the IID realizations of $U_3$.

The parameters $\varsigma_0$ and $\varsigma_1$ correspond to the generators $\Psi_0$ and $\Psi_1$, respectively, with the condition that $\varsigma_0 \leq \varsigma_1$. The parameter $\varsigma_1$ measures the correlation between $U_2$ and $U_3$, while the parameter $\varsigma_0$ measures the correlation between $U_1$ and $U_2$ as well as between $U_1$ and $U_3$, and these two must be equal. This structure is known as nested Archimedean copulas \citep{joe2014dependence}. In this nested formulation, the correlation between the potential death times from each arm is stronger than the correlation between the first event time and the potential death time, which, as discussed, appears more appropriate than assuming the correlation between each pair is the same.

There is no known general multivariate extension of a bivariate family of copulas that has a dependence parameter for each pair of marginals. The normal copula can have three dependence parameters for trivariate copulas, but it is not an extension of the bivariate copula and is challenging to implement. In contrast to Archimedean copulas, the computing time for the normal copula is substantial, as it requires numerical integration without a closed-form solution for each component of the log-likelihood \citep{schemper2013estimating, zimmer2006using}. For our trivariate application, the mixtures-of-powers approach is sufficient, and we will therefore proceed with Archimedean copulas.

\subsubsection{Estimation of Copula Parameters}
Because we can observe $(e^z_i, y^z_i)$ and corresponding status indicators $(\delta_{1i}, \delta_{2i})$ for each subject $i$, we can estimate $\varsigma_0$ for a given copula by maximizing the pseudo-likelihood. The pseudo-likelihood differs from the likelihood function in that $S_E(\cdot)$ and $S_Y(\cdot)$ are estimated nonparametrically using either the Nelson-Aalen or the Kaplan-Meier estimators, instead of specifying a parametric distribution governed by other parameters \citep{joe2014dependence}. The pseudo-likelihood of the copula parameter $\varsigma_0$ is expressed in terms of $\mathbb{C}_{\varsigma_0}(U_1, U_2)$, and its derivative, $\frac{\partial \mathbb{C}_{\varsigma_0}(U_1, U_2)}{\partial U_1}$, $\frac{\partial \mathbb{C}_{\varsigma_0}(U_1, U_2)}{\partial U_2}$, and the second derivative $c_{\varsigma_0}(U_1, U_2)=\partial^2 \mathbb{C}_{\varsigma_0}(U_1, U_2)/\partial U_1 \partial U_2$ as follows \citep{schemper2013estimating}:
\begin{eqnarray}
L(\varsigma_0) &=& \prod_i \Bigg[
    c_{\varsigma_0}(u_{1i}, u_{2i})^{\delta_{1i}\delta_{2i}}  \left\{-\frac{\partial \mathbb{C}_{\varsigma_0}(u_{1i}, u_{2i})}{\partial u_{1i}}\right\}^{\delta_{1i}(1-\delta_{2i})} \nonumber\\[1ex]
&& \quad \times \left\{-\frac{\partial \mathbb{C}_{\varsigma_0}(u_{1i}, u_{2i})}{\partial u_{2i}}\right\}^{(1-\delta_{1i})\delta_{2i}} \mathbb{C}_{\varsigma_0}(u_{1i}, u_{2i})^{(1-\delta_{1i})(1-\delta_{2i})}
\Bigg]. \label{copulalld} 
\end{eqnarray}
 
If baseline covariates are recorded, they can be included by redefining $U_k$, where $k=1,2$, as the conditional survival function. The maximum pseudo-likelihood estimate (MPLE) $\hat \varsigma_0$ is obtained as the value of $\varsigma_0$ that maximizes $L(\varsigma_0)$. 

\subsection{Estimation of Principal Stratum Probabilities}
We defined $\mathcal{S}$ as the set of first non-fatal event times from all subjects, with $m$ event times within $\mathcal{S}$, denoted by $0 < t_1 < t_2 < \dots < t_m$. At each $t_j \in \mathcal{S}$, we defined $\mathcal{R}_{j}$ as the at-risk set at $t_j$, which includes subjects who remain in the study and have not yet experienced a non-fatal event by time $t_j$. Each subject $i$ belonging to $\mathcal{R}_{j}$, whose last known follow-up time is denoted by $D_i=\hbox{min}(Y_i, C_i)$ with realization $d_i$, where $d_i > t_j$, falls into one of the following four cases:
\begin{enumerate}
    \item $Y_i=d_i, E_i=t_j$: The subject died at $d_i$, and the first event occurred at $t_j$.
    \item $Y_i=d_i, E_i>t_j$: The subject died at $d_i$, and the first event has not yet occurred by $t_j$.
    \item $Y_i>d_i, E_i=t_j$: The subject was censored at $d_i$, and the first event occurred at $t_j$.
    \item $Y_i>d_i, E_i>t_j$: The subject was censored at $d_i$, and the first event has not yet occurred by $t_j$.
\end{enumerate}

This is equivalent to the formulation that substitutes $E_i$ with $T_i$, because under the restriction $d_i > t_j$, we have $E_i=\hbox{min}(Y_i, T_i)=T_i$.

Suppose, without loss of generality, that a subject $i$ is from the active treatment arm. If the subject $i$ falls into case 1, the principal stratum probability at $t_j$ is:
\begin{eqnarray*}
    p_{ij}= \frac{P( E^{1}_i=t_j, Y^{1}_i=d_i, Y^{0}_{i}>t_j)}{P(E^{1}_i=t_j, Y^{1}_i=d_i)}=\frac{\partial^2 \mathbb{C}_{\varsigma_0, \varsigma_1}(u_{1i}, u_{2i}, u_{3i})/\partial u_{1i} \partial u_{2i}}{\partial^2 \mathbb{C}_{\varsigma_0}(u_{1i}, u_{2i})/\partial u_{1i} \partial u_{2i}} . 
\end{eqnarray*}

Similarly, if subject $i$ falls into cases 2, 3, and 4, the principal stratum probability at $t_j$ are given by: $\frac{\partial \mathbb{C}_{\varsigma_0, \varsigma_1}(u_{1i}, u_{2i}, u_{3i})/ \partial u_{2i}}{\partial \mathbb{C}_{\varsigma_0}(u_{1i}, u_{2i})/\partial u_{2i}}$, $\frac{\partial \mathbb{C}_{\varsigma_0, \varsigma_1}(u_{1i}, u_{2i}, u_{3i})/\partial u_{1i}}{\partial \mathbb{C}_{\varsigma_0}(u_{1i}, u_{2i})/\partial u_{1i} }$, $\frac{\mathbb{C}_{\varsigma_0, \varsigma_1}(u_{1i}, u_{2i}, u_{3i})}{\mathbb{C}_{\varsigma_0}(u_{1i}, u_{2i})}$, respectively.

The chosen copulas have their $\varsigma_0$ replaced with the MPLE, enabling the computation of the denominator in the principal stratum probabilities formula above. However, $\varsigma_1$ cannot be directly estimated due to the counterfactual nature of $Y_i^0$. To calculate the numerator, a pre-specified value for $\varsigma_1$ is necessary.

\subsection{Estimation of PSHR in the COPERNICUS Trial}
\subsubsection{Visual Assessment of Fit of Copulas}
After calculating the principal stratum probabilities, we can follow the method outlined in the main text to estimate the PSHR. The next step is to select an appropriate copula. To assess the fit of the three copulas to the data, we compare the empirical joint survival $P(E > e_i, Y > y_i)$, for $1\leq i \leq n$, with the joint survival calculated by the empirical marginal survival functions $P(E > e_i)$, $P(Y > y_i)$, and the chosen copula. This comparison focuses on the subgroup of subjects who experienced both the non-fatal event and death, as only in this group are both event times fully observed and ordered, allowing for the empirical estimation of the joint survival at $(e_i, y_i)$ by $\frac{1}{n} \sum_{j=1}^{n} I( e_j > e_i, y_j > y_i)$, where $I(\cdot)$ is the indicator function. Similarly, the survival function of $E$ at $e_i$ can be estimated by $\frac{1}{n}\sum_{j=1}^{n} I(e_j > e_i)$ and the same estimation applies to the empirical survival function of $Y$. 

\begin{figure}[htbp]
    \centering
    \includegraphics[width=1\textwidth]{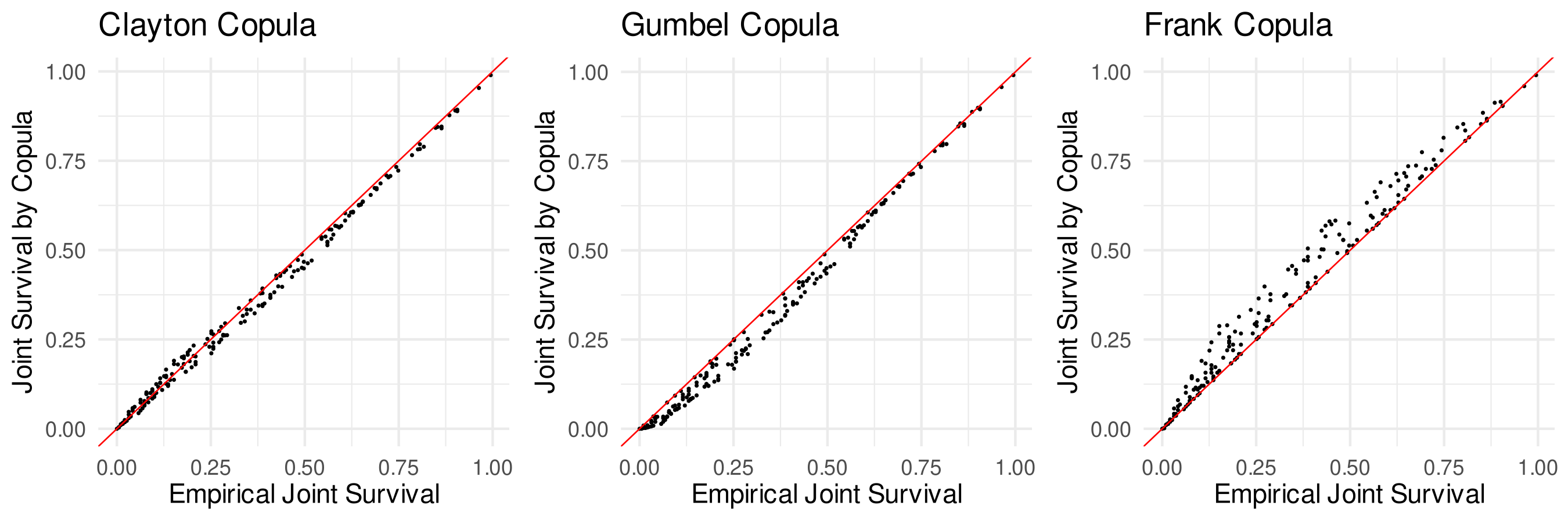}
    \caption{Fit assessment of different copulas for a COPERNICUS trial subgroup.}
    \label{copulacheck}
\end{figure}

In the context of the COPERNICUS trial, without including baseline covariates, the estimates for $\varsigma_0$ for the Clayton, Gumbel, and Frank copulas are 1.11 $(SE=0.13)$, 1.24 $(SE=1.68)$, and 30.23 $(SE=1.92)$, respectively. These estimates are derived by maximizing the pseudo-likelihood in equation (\ref{copulalld}) using $(e_i, y_i)$ and $(\delta_{1i}, \delta_{2i})$ from all subjects. The standard errors are calculated using 200 bootstrap samples. Figure \ref{copulacheck} displays the joint survival estimated by each copula alongside the empirical joint survival, for the 191 subjects who experienced both a non-fatal event and death. For each copula, the parameter $\varsigma_0$ is replaced with its MPLE. The red line represents the reference line where the joint survival estimated by the copula equals the empirical joint survival. It appears that the dots are more closely aligned with the red line when the Clayton copula is used, while the Gumbel copula tends to underestimate the empirical survival and the Frank copula tends to overestimate it. Although the Clayton copula appears to fit well for this subgroup, this does not imply it is the best fit for all subjects. For illustration purposes, we will proceed with the Clayton copula, but the procedure demonstrated above can be applied to other copulas as well.

\subsubsection{Estimation of PSHR Using the Clayton Copula}
With the Clayton copula, the copula functions are defined as $\mathbb{C}_{\varsigma_0, \varsigma_1}(U_1, U_2, U_3)=\{ U_1^{-\varsigma_0} + (U_2^{-\varsigma_1} + U_3^{-\varsigma_1} -1)^{\varsigma_0/\varsigma_1} -1 \}^{-1/\varsigma_0}$ and $\mathbb{C}_{\varsigma_0}(U_1, U_2)=(U_1^{-\varsigma_0} + U_2^{-\varsigma_0} -1)^{-1/\varsigma_0}$. We set $\varsigma_0$ as its MPLE, which is 1.11. Additionally, we vary $\varsigma_1$ across values 1.11, 2, 3, 5, 8, corresponding to Kendall's tau values of 0.36, 0.5, 0.6, 0.71, and 0.8, respectively. Table \ref{coperncopula_mar} presents the estimation results of the PSHR, with the CSHR as the reference. By setting $\varsigma_0=\varsigma_1=0$, we assume mutual independence between $E^z$, $Y^z$, and $Y^{1-z}$, resulting in nearly identical results to the CSHR. When varying $\varsigma_1$ from 1.11 to 8 while fixing $\varsigma_0=1.11$, the estimated PSHR shows only slight variation, ranging from 0.803 to 0.796.

Baseline covariates can be included in the estimation of both $\varsigma_0$ and the PSHR. In addition to treatment assignment, the COPERNICUS trial recorded baseline covariates such as sitting heart rate, history of myocardial infarction, presence of concomitant disease, diabetes, and the use of diuretics, ACE inhibitors, digitalis, amiodarone, and anticoagulants. With these covariates included, $\varsigma_0$ is estimated to be 0.99, 1.23, and 36.31 for the Clayton, Gumbel, and Frank copulas, respectively. Continuing with the Clayton copula, table \ref{coperncopula_cond} presents the estimation results of treatment effects conditional on baseline covariates. Similar to the results in table \ref{coperncopula_mar}, setting $\varsigma_0=\varsigma_1=0$ results in a PSHR virtually identical to the CSHR. By varying $\varsigma_1$ from 0.99 to 8 while fixing $\varsigma_0=0.99$, the estimated PSHR only slightly varies, ranging from 0.790 to 0.784.

\bibliography{supplementary}  

\clearpage

\begin{table}[htbp]
\centering
\caption{Estimation results for the COPERNICUS trial data using trivariate copulas.\label{coperncopula_mar}}
{\tabcolsep=25pt
\begin{tabular}{ccccc}
\toprule
Approach              & $\varsigma_0$ & $\varsigma_1$ & HR & 95\% CI \\ \midrule
CS                  & - & -       & 0.821   & (0.716, 0.943)   \\  \midrule
\multirow{5}{*}{PS} & 0 & 0     & 0.822   & (0.716, 0.948)   \\ 
                    & 1.11 & 1.11      & 0.803   & (0.690, 0.935)  \\
                    & 1.11 & 2      & 0.802   & (0.691, 0.934)  \\
                    & 1.11 & 3      & 0.801   & (0.691, 0.933)  \\
                    & 1.11 & 5      & 0.799   & (0.690, 0.931)  \\
                    & 1.11 & 8     & 0.796  &  (0.688, 0.928)  \\ 
         
\bottomrule
\end{tabular}}
\begin{tablenotes}
\item[] \textbf{Notes:}
\begin{itemize}
\item $\varsigma_0$ and $\varsigma_1$ are the parameters of the Clayton copula. 
\item The hazard ratio (HR) of the PS approach is calculated as $\hbox{exp}(\hat \beta)$.
\item The 95\% CI of the PS approach is derived from 2.5th and 97.5th percentiles of 1000 bootstraps.
\end{itemize}
\end{tablenotes}
\end{table}

\begin{table}[htbp]
\centering
\caption{Estimation results for the COPERNICUS trial data using trivariate copulas, conditioned on baseline covariates. \label{coperncopula_cond}}
{\tabcolsep=25pt
\begin{tabular}{ccccc}
\toprule
Approach              & $\varsigma_0$ & $\varsigma_1$ & HR & 95\% CI \\ \midrule
CS                  & - & -       & 0.809   & (0.704, 0.929)   \\  \midrule
\multirow{5}{*}{PS} & 0 & 0     & 0.809   & (0.693, 0.933)   \\ 
                    & 0.99 & 0.99      & 0.790   & (0.674, 0.920)  \\
                    & 0.99 & 2      & 0.789   & (0.675, 0.919)  \\
                    & 0.99 & 3      & 0.788   & (0.675, 0.919)  \\
                    & 0.99 & 5      & 0.786   & (0.672, 0.917)  \\
                    & 0.99 & 8     & 0.784  &  (0.668, 0.915)  \\ 
         
\bottomrule
\end{tabular}}
\begin{tablenotes}
\item[] \textbf{Notes:}
\begin{itemize}
\item Baseline covariates include:
\begin{itemize}
    \item Sitting heart rate
    \item History of myocardial infarction
    \item Presence of concomitant disease
    \item Diabetes
    \item Use of diuretics
    \item Use of ACE inhibitors
    \item Use of digitalis
    \item Use of amiodarone
    \item Use of anticoagulants
\end{itemize}
\end{itemize}
\end{tablenotes}
\end{table}